\documentclass[final,3p,times,twocolumn]{elsarticle}
\usepackage{amssymb}
\usepackage{soul}
\usepackage{amsmath}
\usepackage{float}
\usepackage{xcolor}
\definecolor{RED}{RGB}{156,78,90}

\journal{Current Opinion in Colloid $\&$ Interface Science}

\usepackage{lineno}
\usepackage{setspace}
\usepackage{xcolor}

\usepackage{tabularx}
\usepackage{makecell} 
\usepackage{ragged2e} 
\usepackage{booktabs} 

\usepackage{color}
\definecolor{B}{RGB}{52,78,200}

\begin{document}
\begin{sloppypar}
\begin{frontmatter}

\title{Recent advances in stimulus-assisted nanoprecipitation for nanoparticle synthesis}

\author[1,2]{Mingbo Li\corref{cor1}}
\ead{mingboli@sjtu.edu.cn}
\cortext[cor1]{Corresponding author}
\address[1]{School of Ocean and Civil Engineering, Shanghai Jiao Tong University, Shanghai 200240, China}
\address[2]{Key Laboratory of Hydrodynamics (Ministry of Education), Shanghai Jiao Tong University, Shanghai 200240, China}

\author[1]{Junhao Cai}

\author[3]{Yawen Gao\corref{cor1}}
\ead{yawengao@mail.tsinghua.edu.cn}
\address[3]{New Cornerstone Science Laboratory, Center for Combustion Energy, Key Laboratory for Thermal Science and Power Engineering of Ministry of Education, Department of Energy and Power Engineering, Tsinghua University, Beijing 100084, China}


\begin{abstract}
Nanoprecipitation, the rapid solvent-displacement route to nanoscale phase separation, has matured from a simple batch operation into a versatile platform for nanomaterial synthesis. This review synthesizes recent progress in stimulus-assisted nanoprecipitation, wherein externally applied triggers (ultrasonic, electrical, supergravity, thermal, chemical, and photonic/other stimuli) are integrated with contemporary mixing technologies (batch, flash, microfluidic, membrane and high-shear reactors) to decouple and selectively control over nucleation, growth kinetics, and assembly processes. These methods allow for the precise tuning of the size, morphology, stability and functionality of nanoparticles (NPs), thereby broadening their applications in drug delivery, catalysis and materials science. We distill mechanistic principles by which each stimulus alters local supersaturation, chain mobility, interfacial instabilities, or droplet/film microreactor dynamics, and compare advantages and limitations by surveying research works from recent years. We also explore the potential development trends of multiscale coupling models, design rules for stimulus-compatible continuous reactors, and adoption of data-driven optimization frameworks to expand the capabilities of nanoprecipitation for advanced nanomaterial design.
\end{abstract}


\end{frontmatter}


\section{Introduction} 

Nanoprecipitation ("Ouzo effect", solvent shifting, or anti‐solvent precipitation) is a versatile and low-energy method for producing nanoparticles (NPs) by mixing a solute solution (e.g., a polymer or small molecule) with a miscible anti-solvent (e.g., water or buffer solution)~\cite{yan2021nanoprecipitation}. Rapid solvent exchange induces supersaturation, chain collapse, and phase separation into NPs spanning from a few nanometers to several micrometers~\cite{kuddushi2025recent}. Depending on supersaturation, nuclei either grow gradually ("nucleation-growth") or aggregate ("nucleation-aggregation"), yielding NPs with different sizes and dispersities. Rapid mixing generally favors small/monodisperse NPs, while slower mixing produces larger/polydisperse NPs. Recent work~\cite{fesenmeier2024overcoming} showed that Equilibration-Nanoprecipitation, which equilibrates micelles before solvent removal, enhances monodispersity compared to conventional methods.  

Batch nanoprecipitation, achieved by one-step or dropwise addition of solute solution into anti-solvent, remains the simplest approach. Continuous techniques such as flash nanoprecipitation (FNP) and microfluidic nanoprecipitation allow ultrafast and reproducible mixing within milliseconds. Despite differences in mixing, both rely on the solvent-displacement mechanism and are applicable to hydrophobic and hydrophilic solutes~\cite{kuddushi2025recent}. To expand structural control, sequential nanoprecipitation separates core formation from stabilization. This decoupling enables independent regulation of aggregation kinetics and surface coverage, which is difficult in single-step mixing. Programmable nanoprecipitation with multi-step solvent shifting further allows on-demand engineering of polymer NPs with well-defined morphologies~\cite{liu2020stable, yan2019programmable}. Up to now, conventional~\cite{liu2020stable}, microfluidic~\cite{li2025multistage}, and flash~\cite{fu2020direct} approaches all have been adapted to sequential nanoprecipitation to generate NPs with tailored structurings~\cite{lewis2025process}.   

Nanoprecipitation has enabled a wide spectrum of nanomaterials, including polymer NPs, cellulose acetate NPs, food-grade NPs, lipid NPs, semiconductor nanocrystals, Janus NPs, and mesoporous NPs~\cite{yan2021nanoprecipitation, chen2023recent}. Consequently, diverse nanostructures such as compact, core–shell, porous, Janus, hollow, trilobal, lamellar, patchy, and Janus–lamellar morphologies have been obtained~\cite{chen2023recent, sun2021diverse}. The resulting properties are strongly influenced by solvent composition, mixing rate, stabilizers, and physicochemical parameters. Amphiphilic copolymers, for example, self-assemble into micelles, vesicles, liposomes, polymersomes, or fibers during solvent shifting, often achieving stability over days or months without surfactants~\cite{sharratt2021precision}. 

Particularly, external stimuli have been integrated into nanoprecipitation to modulate NP size, morphology, stability, and functionality. These stimulus-assisted approaches extend the technique beyond solvent mixing, broadening its utility for nanomaterials in drug delivery, catalysis, optics, and related applications. In this review, we first summarize recent advances in mixing strategies, then focus on six classes of stimulus-assisted nanoprecipitation, highlighting mechanistic principles, advantages and limitations, and representative examples. Our goal is to provide a comprehensive yet concise overview of how stimuli have been leveraged to expand the nanoprecipitation toolbox and to inspire further developments in nanomaterial design.  

\section{Rapid mixing methods for controlling nanoprecipitation}

\begin{figure*}[!t]
\centering
\includegraphics[width = 0.97\textwidth]{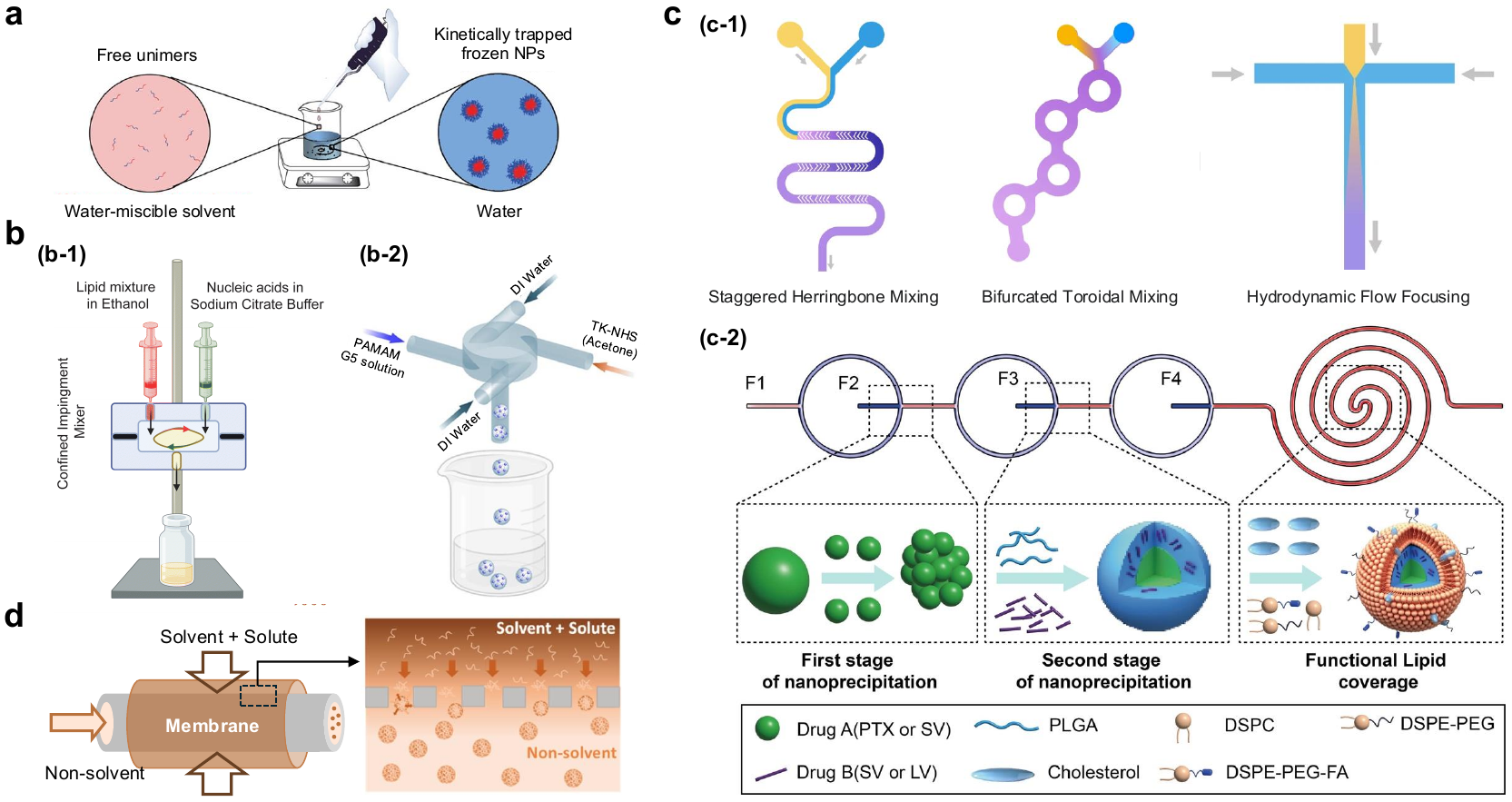}
\caption{Typical mixing approaches for nanoprecipitation. 
(a) Conventional batch nanoprecipitation. Adapted from Ref.~\cite{bovone2022solvent}. 
(b) FNP methods. (b-1) A confined impingement jet mixer (CIJ); (b-2) a multi-inlet vortex mixer (MIVM). Adapted from Refs.~\cite{misra2024flash, qi2025synthesis}. 
(c) Microfluidic nanoprecipitation. (c-1) Representative schematics for microfluidic mixing geometries; (c-2) schematic of multi-stage microfluidic assisted co-delivery platform for multi-agent facile synchronous encapsulation. Adapted from Refs.~\cite{vandenberg2025learning, li2025multistage}. 
(d) Membrane nanoprecipitation. Adapted from Ref.~\cite{piacentini2022membrane}.
}
\label{FIG1}
\end{figure*}

Different rapid-mixing approaches, conventional batch, flash, microfluidic, membrane and a range of alternative reactors, provide distinct and complementary levers to control nanoprecipitation outcomes (size, dispersity, morphology) while addressing scalability and reproducibility challenges~\cite{kuddushi2025recent}. This section briefly surveys these mixing approaches. In conventional batch nanoprecipitation (FIG.~\ref{FIG1}(a)), typically, the antisolvent is typically introduced to a stirred tank by dropwise addition, pump dosing, or by diffusion across dialysis membranes~\cite{bovone2022solvent}. The strengths of batch methods are operational simplicity and reproducibility, but they are limited in achieving the extreme temporal and spatial mixing control provided by continuous technologies.

FNP exploits millisecond-scale, turbulence-enhanced mixing within confined geometries to produce narrowly distributed NPs. Devices such as confined impingement jets (CIJ) and multi-inlet vortex mixers (MIVM) (FIG.~\ref{FIG1}(b-1) and (b-2)) generate intense micromixing that favors formation of numerous small nuclei and suppresses broad secondary growth~\cite{misra2024flash, qi2025synthesis}. Hardware and chemistry innovations, e.g., droplet evaporation approaches without high-speed jets~\cite{manohar2024drying} or in-flight crosslinking during flash mixing~\cite{qi2025synthesis}, broaden the palette of attainable nanostructures and payload chemistries, illustrating how mixing geometry and solvent dynamics jointly determine outcomes. 

Microfluidic nanoprecipitation translates precise channel geometry and flow control into highly reproducible, tunable NP synthesis. Passive microfluidic mixers accelerate mixing under laminar flow by reducing diffusion paths and leveraging chaotic advection; commonly used geometries include the staggered-herringbone and bifurcated toroidal mixers (FIG.~\ref{FIG1}(c-1)). Platforms range from hydrodynamic flow focusers and droplet reactors to channels with embedded baffles or microstructures~\cite{vandenberg2025learning}. Microfluidic designs excel at sequential or multistage mixing, enabling programmed solvent shifts and complex assembly pathways~\cite{li2025multistage} (FIG.~\ref{FIG1}(c-2)).

Membrane nanoprecipitation exploits porous substrates to generate local supersaturation at pore outlets, enabling continuous production and alleviating batch limitations (FIG.~\ref{FIG1}(d)). Experimental evidence indicates that solvent/non-solvent volumetric ratio often dominates particle formation behavior in membrane systems, more so than transmembrane flux or shear, supporting continuous, scalable operation with relatively mild hydrodynamic stresses~\cite{piacentini2022membrane}. Despite remaining gaps in mechanistic understanding of the coupled chemical–membrane–fluid dynamics, membrane-assisted approaches show promise for high-throughput manufacture.

Beyond these mainstream routes, alternative reactors, such as spinning disc systems (SDR)~\cite{mens2024spinning}, evaporation-triggered spinning-disc processes~\cite{zander2025evaporation}, and inline high-shear mixers (HSM)~\cite{tang2025research}, offer additional pathways to rapid solvent removal and confined growth. SDR systems form thin films with high surface-area-to-volume ratios that limit nucleus collisions and yield smaller, more uniform NPs. HSM approaches couple intense micromixing with shear-driven deagglomeration to produce NPs in continuous, clogging-resistant processes, with predictive scale-up models and electrostatic stabilization strategies~\cite{tang2025research}. New mixing modes and process intensification equipment are emerging~\cite{feng2025microchannels}, selection among methods therefore depends on the required temporal resolution, desired architecture (e.g., core–shell, cross-linked), payload chemistry, and scale-up constraints.

\section{External stimulus-assisted nanoprecipitation}

\begin{figure}[!t]
\centering
\includegraphics[width = 0.45\textwidth]{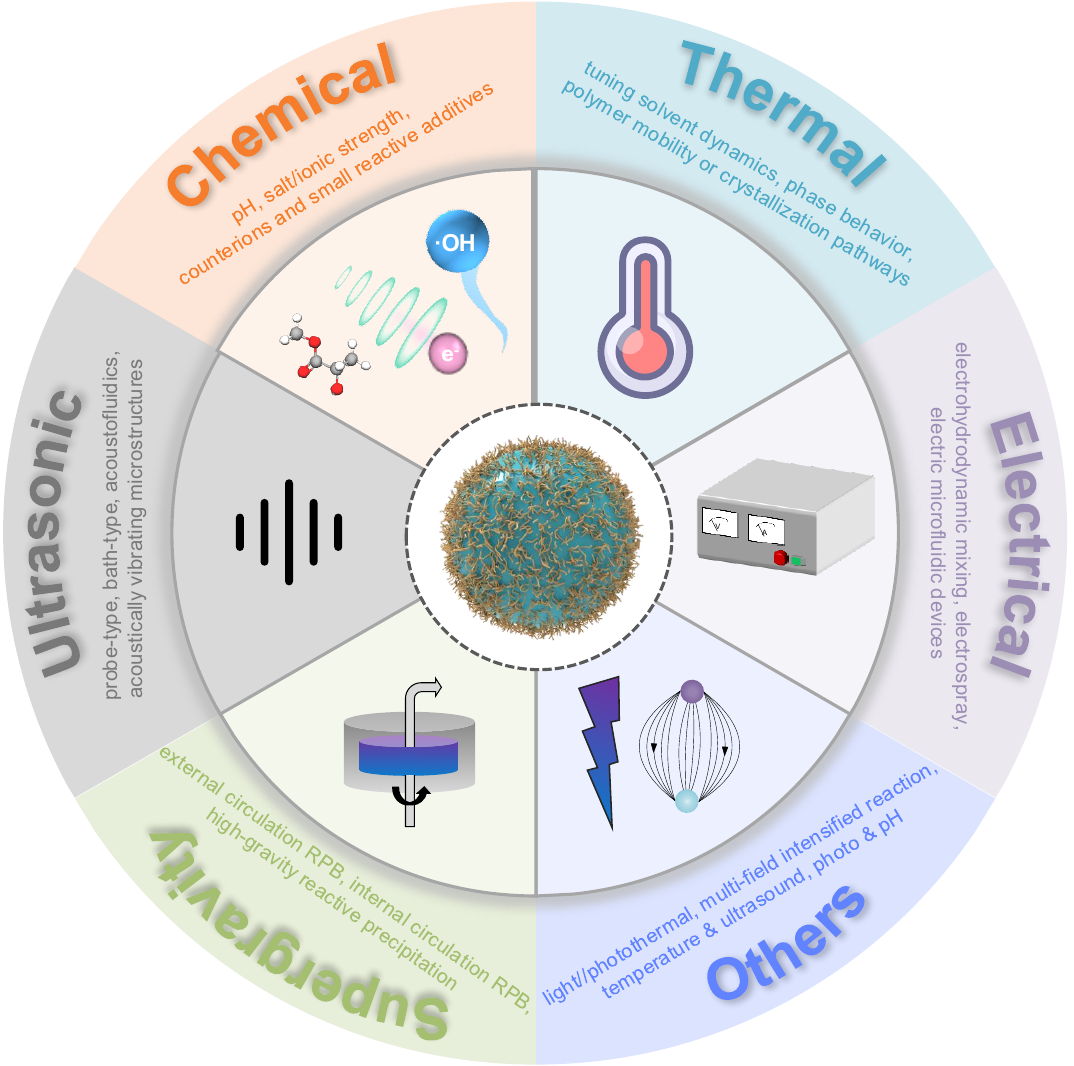}
\caption{Schematic illustration of different types of external stimuli for nanoprecipitation. }
\label{FIG2}
\end{figure}

Based on the above mixing process, external stimuli modulate the two kinetic axes that determine nanoprecipitation outcomes: 1) Nucleation kinetics (rate and number of nuclei); 2) Growth and reorganization kinetics (chain mobility, crystallization, or coalescence). Controlling these governs size distributions, polydispersity, self-assembled morphology, or crystallinity/amorphous content. We next examine six major categories of factors (FIG.~\ref{FIG2}), including ultrasonic-, electrical-, supergravity-, thermal-, chemical-, and other external stimuli, discussing the underlying mechanisms and recent experimental examples in material synthesis and interface engineering.

\subsection{Ultrasonic-assisted nanoprecipitation}

Ultrasound (typically 20 kHz$\sim$2 MHz) dramatically enhances nanoprecipitation by converting acoustic energy into intense local hydrodynamic and thermodynamic perturbations via cavitation. In ultrasound-assisted nanoprecipitation, the nucleation and growth pathways are reshaped: cavitation bubbles form, grow and collapse, producing localized high temperatures, transient high-pressures, shockwaves and microjets that generate extreme microscale shear and impulsive mixing. These effects produce nearly instantaneous, spatially uniform supersaturation and thereby favor rapid, homogeneous nucleation over diffusion-limited growth~\cite{feng2025microchannels}. Complementary mechanisms (acoustic microstreaming and steady circulatory flows) enhance bulk mass transfer, thin diffusion layers and homogenize stabilizer distribution, while modest bulk heating lowers viscosity and accelerates molecular diffusion. These phenomena shift precipitation kinetics from growth-dominated to mixing- and nucleation-dominated regimes, suppressing NP growth and aggregation~\cite{sharma2019effect}. Ultrasonic shear also deagglomerates nascent clusters, facilitating uniform stabilizer adsorption and often increasing measured zeta potential, which improves colloidal stability~\cite{benhabiles2025performance}. 

The main types of ultrasound-assisted nanoprecipitation include: probe-type, bath-type, acoustic micromixers and combined methods. Each approach exploits the same physical effects (cavitation and microstreaming) but differs in frequency, scale, throughput, and energy efficiency~\cite{huang2019acoustofluidic}. For example, probe sonication provides more intense cavitation locally (lab-scale syntheses)~\cite{zhang2024physicochemical}, whereas bath sonication is milder but can be used to agitate larger volumes uniformly~\cite{wang2022formation}. For example, synchronized use of a high-speed homogenizer and a probe sonicator has been reported to produce ultrafine flavonoid particles via anti-solvent precipitation~\cite{zhang2024antisolvent}. Although the batch reactor model is simple to operate, it face limitations due to uneven acoustic field distribution and challenges in scalability.

Acoustics-integrated microfluidics (acoustofluidics) provide a powerful platform for controlled nanoprecipitation by combining precise fluid handling with acoustic energy to accelerate mixing, enhance mass transfer, and tailor nucleation kinetics. These devices, classified broadly as bulk acoustic waves (BAWs), surface acoustic waves (SAWs) and acoustically vibrating microstructures, offer distinct mechanisms to generate microvortices and shear at low power, yielding advantages in biocompatibility, simplicity and cost-effectiveness~\cite{kamat2023active}. 

\begin{figure*}[!t]
\centering
\includegraphics[width = 0.98\textwidth]{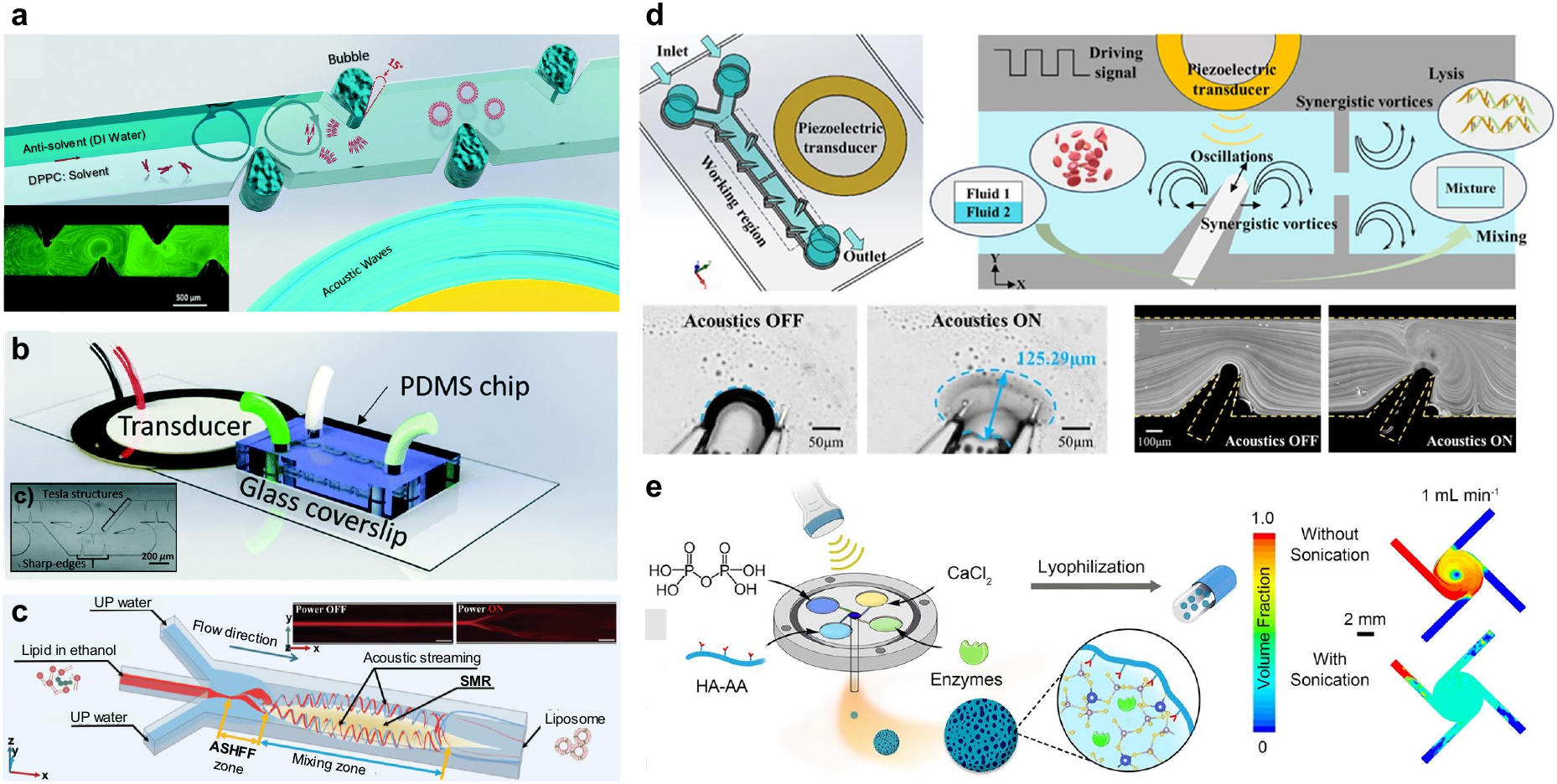}
\caption{(a) Microbubble-integrated sidewall wedge acoustic micromixer. The inset: fluorescent polystyrene particle behavior in the presence of an acoustic field. Adapted from Ref.~\cite{rasouli2019ultra}. (b) Tesla structure-integrated sidewall wedge acoustofluidic mixer. Adapted from Ref.~\cite{bachman2020acoustofluidic}.
(c) Schematic of the liposome synthesis platform based on an solid mount resonator (SMR). Adapted from Ref.~\cite{xu2024microfluidic}.
(d) Top panel: schematic and working mechanism of the acousto-inertial chip. Bottom panel: gas-liquid interface deformation of an oscillating bubble with acoustic excitation (left); fluorescent polystyrene particle tracing the microvortex streaming (right). Adapted from Ref.~\cite{lu2024vortex}.
(e) Synthesis of CaP-enzyme nanocatalysts with a multi-inlet acoustofluidic mixer. Adapted from Ref.~\cite{wu2025acoustofluidic}.  
}
\label{FIG3}
\end{figure*}

Among of them, vibrating microstructures (sharp edges, micropillars, trapped bubbles) produce strong acoustic streaming and have therefore attracted substantial attention~\cite{chen2021sharp}. Sharp edges act as robust, reproducible boundaries that convert pressure oscillations into steady microstreams, enabling efficient mixing and pumping without the instability associated with trapped bubbles. Bubble-based acoustofluidics, while delivering intense local streaming, is often constrained by bubble-trapping complexity and structural instability during operation~\cite{ozcelik2014acoustofluidic}. Hybrid designs that combine bubbles and solid microfeatures can capture the energetic benefits of both while mitigating individual drawbacks. Rasouli and Tabrizian~\cite{rasouli2019ultra} proposed a micromixer integrating oscillating bubbles with sharp-edge structures to generate ultra-rapid, boundary-driven microvortices (FIG.~\ref{FIG3}(a)), and enabling precise nanoprecipitation of organic NPs compared to hydrodynamic flow focusing~\cite{agha2024integration}. Bachman et al.~\cite{bachman2020acoustofluidic} combined sharp-edge acoustic streaming with passive Tesla-structured hydrodynamics in a single-layer PDMS device, merging active mixing at low throughput with passive vortexing at higher flow rates to deliver consistent PLGA–PEG NP synthesis across variable throughputs (FIG.~\ref{FIG3}(b)). Such hybridization improves robustness but remains susceptible to clogging and dilution limitations inherent to microstructured channels. Diverging from kHz sharp-edge strategies, GHz-frequency BAW resonators (e.g., SFLS-SMR) exploit high-frequency energy confinement and rapid dissipation to drive body-force-driven microvortices without complex microfeatures (FIG.~\ref{FIG3}(c)). By eliminating micro-obstacles, this approach reduces clogging risk and realizes flow-rate–independent tuning of liposome size through acoustic power modulation alone, simplifying operation and control~\cite{xu2024microfluidic}. Complementary advances, such as acousto-inertial chips~\cite{lu2024vortex} that integrate bubble–sharp-edge composites with contraction–expansion arrays, capitalize on synergistic acoustic and inertial forces to produce stable, high-intensity vortices across broad flow ranges at low excitation voltages (FIG.~\ref{FIG3}(d)). Beyond hydrodynamic mixing, acoustofluidic cavitation has been integrated into FNP. A multi-inlet acoustofluidic mixer~\cite{wu2025acoustofluidic} recently demonstrated ultrafast, homogeneous mixing via combined acoustic cavitation and streaming, enabling scalable synthesis of calcium pyrophosphate–enzyme nanocatalysts by FNP (FIG.~\ref{FIG3}(e)) and pointing to new ultrasound-assisted routes for rapid NP formation. 

By judiciously selecting acoustic modality and combining active and passive elements, researchers can precisely tune local mixing timescales and nucleation rates to control particle size, dispersity and morphology. A broad range of organic, inorganic and hybrid materials has been produced in acoustofluidic reactors~\cite{rasouli2023acoustofluidics}. Moreover, ultrasonically generated NPs frequently display porous morphologies and high surface areas, which enhance adsorption properties and catalytic activity across applications~\cite{buyukkanber2024high}. Scale-up, however, is constrained by the energetics and limited acoustic penetration of cavitation: high-intensity ultrasound is power-intensive and attenuates in bulk fluids, so reactor geometry and transducer placement are critical~\cite{rasouli2023acoustofluidics}. Promising strategies include megasonic or multi-frequency excitation, embedded transducers in continuous-flow reactors, and integration of real-time particle monitoring with closed-loop control to reduce cavitation damage and ensure uniform acoustic fields.

\subsection{Electrical-assisted nanoprecipitation} 

\begin{figure*}[!t]
\centering
\includegraphics[width = 0.98\textwidth]{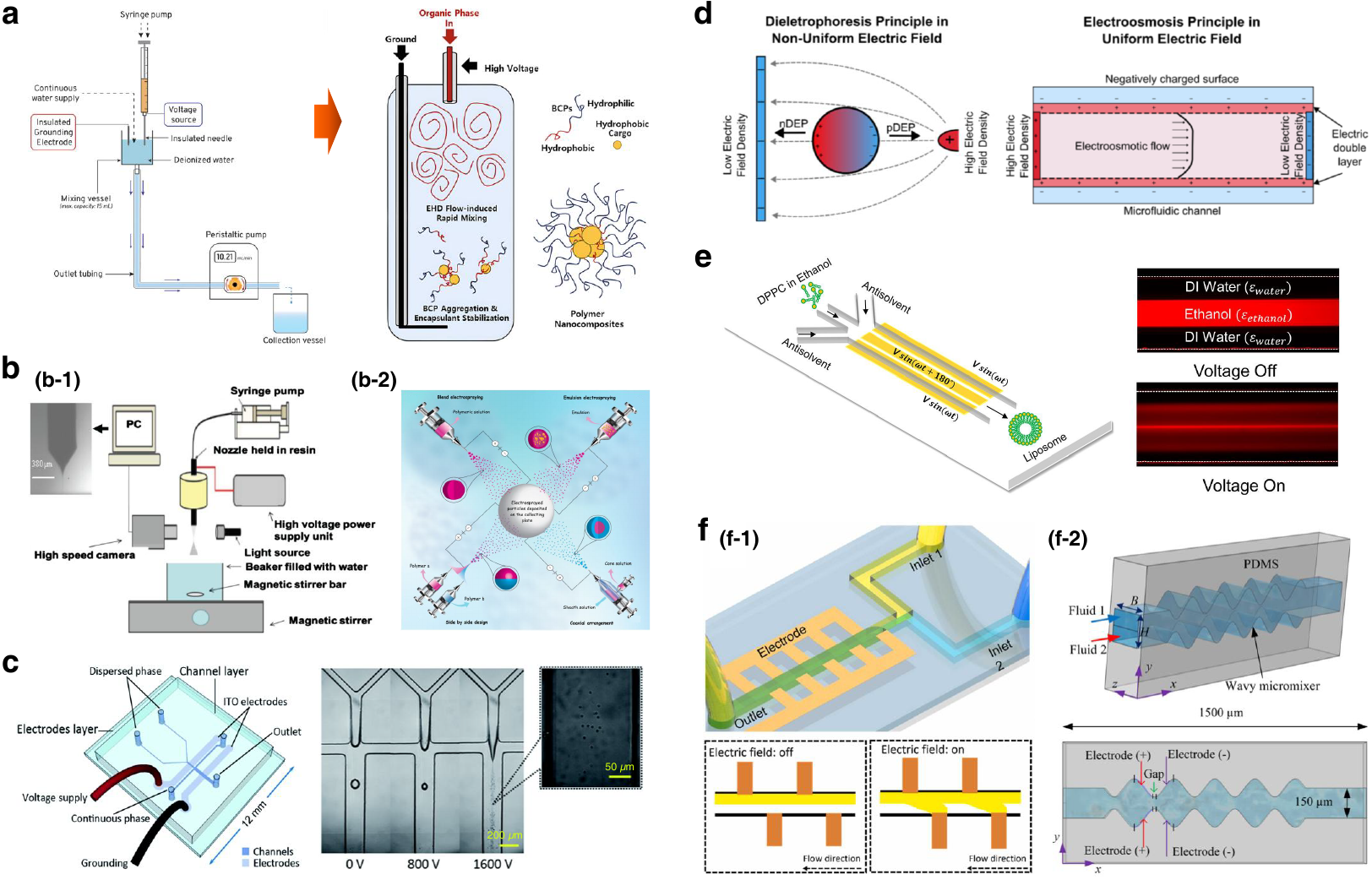}
\caption{Electrical-assisted nanoprecipitation. (a) Schematic of the continuous electrohydrodynamic mixing nanoprecipitation apparatus. Adapted from Ref.~\cite{lee2024semibatch}. 
(b) Electrospraying technique. (b-1) Schematic representation of the electrospray nanoprecipitation set-up; (b-2) Schematic illustration of different set-ups for electrospraying. Adapted from Refs.~\cite{luo2015preparation, rostamabadi2021electrospraying}. 
(c) Schematic of the electrospray microfluidic chip and snapshots of the droplets generated under different voltages. Adapted from Ref.~\cite{sun2016controlled}. 
(d) Dielectrophoresis and electroosmotic flow principles. Adapted from Ref.~\cite{wu2025design}. 
(e) Schematic illustration of the micromixer for NP synthesis using AC electroosmosis, and the fluorescent images showing laminar streams of DI water/ethanol when the voltage is off and on. Adapted from Ref.~\cite{modarres2020electrohydrodynamic}. 
(f) Electric microfluidic devices. (f-1) AC electroosmosis micromixing on a lab-on-a-foil electric microfluidic device; (f-2) AC electrothermal wavy micromixer. Adapted from Refs.~\cite{wu2022ac, mehta2023ac}.
}
\label{FIG4}
\end{figure*}

In electrical-assisted nanoprecipitation, externally applied electric fields or currents are used to enhance or control mixing and precipitation. Two primary modes are electrohydrodynamic (EHD) mixing and electrospray-assisted nanoprecipitation. In EHD mixing, an electric field (DC/AC) is applied to fluid streams or electrodes in contact with the solvents. The field drives fluid motion (by electroosmotic or electrothermal flows) and charge-induced instabilities, generating intense micromixing and rapid solvent displacement. Recent work~\cite{lee2024semibatch} presents that, a continuous electrohydrodynamic mixing-mediated nanoprecipitation system, which maintains a constant organic-to-aqueous ratio, can improves reproducibility, and enhances encapsulation efficiency compared to the semi-batch process (FIG.~\ref{FIG4}(a)). Compared to conventional FNP, EHD mixing can achieve turbulent mixing at much lower flow rates. This means that nanoprecipitation occurs under far-from-equilibrium, kinetically-driven conditions. The charged solvent interface "jets" into the antisolvent, and charge repulsion can further break up droplets.

In electrospray-assisted nanoprecipitation, a polymer solution is fed through a high-voltage nozzle to form a charged liquid jet that disintegrates into fine, highly charged droplets; these droplets mix with antisolvent and evaporate rapidly, acting as individual mini-precipitation vessels that yield monodisperse, sub-micron NPs~\cite{luo2015preparation} (FIG.~\ref{FIG4}(b-1)). Polymer chains collapse within each droplet as solvent is removed or diluted, producing single NPs with high encapsulation efficiency in many cases~\cite{rostamabadi2021electrospraying}. Product morphology and size are highly sensitive to operating parameters, such as electric-field configuration (voltage, electrode spacing/geometry), solution properties, and flow rate, which complicates reproducibility and process validation~\cite{roshan2024curcumin}. Despite its ability to generate structurally complex or compartmentalized NPs, electrospray nanoprecipitation remains limited in practical uptake. The method relies on low-flow, point-source regimes that are difficult to scale without precise multinozzle parallelization, and charged-droplet phenomena (Coulomb fission, secondary breakup) broaden size distributions. Consequently, electrospray and nanoprecipitation are often used sequentially to construct complex nano-micro systems~\cite{zhang2022nano}, and specialized electrospraying variants have been developed within electrically particulated platforms~\cite{rostamabadi2021electrospraying} (FIG.~\ref{FIG4}(b-2)). Integration with microfluidics represents a key experimental advance (FIG.~\ref{FIG4}(c)): on-chip electrospray enables precise generation of Janus or compartmentalized droplets that, upon rapid solvent diffusion into a non-solvent, solidify into uniform Janus NPs~\cite{sun2016controlled}. However, on-chip deposition, fluidic compatibility, and dynamic stability remain technical challenge. Consequently, techniques like FNP are preferred for scalable manufacturing, while electrospray is specialized for architecturally complex NPs~\cite{wang2025electrosprayed}.

Embedded electrodes in microfluidic devices offer a practical and versatile means to actively control mixing and particle assembly via electrokinetic forcing. Integrating electrodes introduces two principal phenomena: dielectrophoresis and electroosmotic flow, which, when engineered through electrode geometry and fabrication, enable precise manipulation of fluids and suspended species~\cite{wu2025design} (FIG.~\ref{FIG4}(d)). Modarres and Tabrizian~\cite{modarres2020electrohydrodynamic} exploited these effects in an electrohydrodynamic micromixer for nanoprecipitation of highly monodisperse liposomes: transverse electric fields imposed by embedded electrodes at the solvent/antisolvent interface drive fluid motion due to abrupt electrical-parameter discontinuities, enabling mixing at low AC voltages (FIG.~\ref{FIG4}(e)). Extending this concept, their phase-controlled field-effect micromixer uses three-finger sinusoidal electrodes whose source-gate-drain phase relations switch the device between unmixed and mixed states, offering temporal programmability of mixing~\cite{modarres2020phase}.
Planar, tooth-shaped electrodes on flexible lab-on-foil substrates similarly generate AC electroosmotic vortices perpendicular to flow, achieving efficient mixing at low frequencies and voltages~\cite{wu2022ac} (FIG.~\ref{FIG4}(f-1)). At higher frequencies, beyond the electrolyte charge-relaxation time, AC electrothermal forces dominate: localized Joule heating from embedded electrodes produces temperature and conductivity gradients that drive strong three-dimensional electrothermal vortices, as implemented in wavy-channel mixers for biofluid applications~\cite{mehta2023ac} (FIG.~\ref{FIG4}(f-2)). This mechanism is well suited to lab-on-a-chip diagnostics, biosensing, and biochemical assays. Collectively, design considerations hinge on electrode patterning, material compatibility and the electrical properties of working fluids, which determine whether electroosmotic, dielectrophoretic or electrothermal mechanisms predominate. 

In this context, electrical-assisted nanoprecipitation offer a viable approach for the synthesis of a wide variety of NPs~\cite{lee2021polymer, chatterjee2020electrospray} (polymer drug nanocarriers, fluorescent polymer NPs, semiconductor NPs and so on), operating with low voltages and without moving parts or additional chemicals. Moreover, this nanoprecipitation technique typically uses deionised water and solvents with low conductivity (e.g. ethanol, isopropanol and acetonitrile), which are ideal for electrical fluid actuation mechanisms as Faradaic reactions and product contamination are highly unlikely.

\subsection{Supergravity-assisted nanoprecipitation}

Supergravity (rotating packed bed, RPB) reactors intensify mixing, mass transfer, and reactions by subjecting fluids to a high-gravity field (1$\sim$3 orders of magnitude greater than $g$)~\cite{chen2000synthesis}. Under these conditions the inlet liquid is rapidly atomized into droplets, ligaments, and thin films across porous packing, producing ultra-fast micromixing and vastly enhanced interfacial area. The resulting high supersaturation triggers a rapid “burst nucleation” that yields small, uniform NPs and shortens reaction times~\cite{tao2024highly}. Estimated micro-mixing times in RPBs are 0.01$\sim$0.1 ms, substantially shorter than typical nucleation induction times ($\sim$1 ms) in aqueous systems, so nucleation occurs within confined microreactors that curtail subsequent growth. Two dominant flow patterns (ligament and droplet) and two disintegration modes (film-droplet and film-ligament–droplet) are commonly observed on the outer packing layers, governing NP formation dynamics~\cite{liu2019cfd}. RPB enables continuous operation and straightforward scale-up, making supergravity nanoprecipitation suitable for ultrasmall, monodisperse NP production. There are two types of RPB reactors widely used: external circulation RPB (ECRPB) and internal circulation RPB (ICRPB) reactors (FIG.~\ref{FIG5}). A schematic of the high-gravity synthesis using ECRPB is shown in FIG.~\ref{FIG5}(a-1). Intensified micro-mixing and mass transfer in the high-gravity field create myriad droplet/film/wire microreactors where metal salts and ligands mix rapidly, driving local concentrations above nucleation thresholds and terminating growth by precursor depletion; the result is highly monodisperse metal–organic framework NPs (MOFs) with sub-5 nm sizes, even close to the length of one crystal unit cell~\cite{chang2021general}. In comparison, ECRPB uses recirculation loops and residence times that favor instantaneous, nucleation-dominated processes but limit control of slower kinetics~\cite{fang2021preparation, chang2021general}, while ICRPB retains fluid in the rotor, allowing adjustable residence times and high-$G$ exposure for balanced nucleation and growth~\cite{guo2023universal}. 

The fundamental distinction between ECRPB and ICRPB reactors lies in their ability to engineer the residence time distribution (RTD) and the shear environment, thereby controlling the kinetic competition between nucleation and growth. In an ECRPB, the reaction mixture circulates between an external shell and the rotating packing. This design often results in a broader RTD, where a portion of the fluid elements experiences longer residence times. While this facilitates overall mass transfer, it can lead to premature growth of initially formed nuclei before the entire solution has passed through the high-shear zone, potentially yielding a wider particle size distribution. Conversely, the ICRPB is ingeniously designed to confine the entire fluid path within the rotor itself, often incorporating internal structures that force a more uniform and typically shorter, narrower RTD (FIG.~\ref{FIG5}(a-2)). This ensures that nearly all precursor molecules are subjected to an intense, uniform shear field almost simultaneously. This rapid and homogeneous exposure achieves an instantaneous, system-wide supersaturation that maximizes "burst nucleation" according to the LaMer model. By rapidly consuming monomers in this initial nucleation burst and minimizing the time for subsequent growth, the ICRPB provides superior control in decoupling nucleation from growth, thereby favoring the dominance of nucleation and the consistent production of small, monodisperse NPs.

This technology has been successfully employed to prepare a variety of NPs~\cite{dong2024instantaneous, zhang2024preparation, wang2022surfactant} (inorganic/organic, pesticide, MOFs) with small sizes and narrow distributions, suitable for a wide range of applications. Recently the integration of high gravity technology with in situ modification for synthesis opens new avenues to produce monodispersed and stable NPs (see the recent review paper~\cite{wang2025synthesis}). This method effectively avoids potential secondary agglomeration during subsequent processing stages, notably enhancing the dispersibility and stability of NPs. The main drawback is specialized equipment and cost. High-speed rotors and sealed housings are needed to contain the rotating flow; maintenance of bearings and packing becomes an issue.

\subsection{Thermal-assisted nanoprecipitation}

Temperature-assisted nanoprecipitation refines conventional solvent-shift methods by deliberately controlling the temperature of one or both phases during mixing or in downstream processing, thereby tuning solvent dynamics, phase behavior, polymer mobility and crystallization pathways~\cite{li2023thermal}. Under these conditions two synergistic effects accelerate NP formation. First, elevated temperature enhances solvent diffusion and evaporation, when the antisolvent temperature exceeds the organic solvent boiling point, rapid solvent removal induces sharp supersaturation spikes that kinetically trap polymer into nuclei~\cite{zhou2017polymeric}. Second, temperature modifies polymer chain mobility and phase behavior: heating above a hydrophobic block’s melting point suppresses premature crystallization in semi-crystalline copolymers, while cooling can drive coil-to-globule collapse, both serving as thermal “triggers” for assembly~\cite{kankanen2020evaluation}. Drying conditions further influence solid-state structure, as shown for amylose NPs whose V-type crystallinity develops during drying and is sensitive to temperature and humidity~\cite{yan2017effect}. 

Recent work has extended temperature-assisted nanoprecipitation across a diverse range of nanosystems, including polymeric vesicles and micelles~\cite{zhou2017polymeric, huang2018tuning}, lipid–polymer hybrids~\cite{wang2020lipidation}, polyanhydride/drug composites~\cite{tao2019application} and other polymeric NPs~\cite{perumal2022review, zander2025evaporation}. Temperature alters both solvent transport and polymer thermodynamics, thereby modulating supersaturation kinetics, nucleation rates and subsequent growth or reorganization. Early studies in FNP demonstrated that lower temperatures reduce equilibrium solubility, elevate supersaturation and hence increase nucleation while suppressing coagulation and Ostwald ripening, motivating ice-bath protocols to yield smaller, narrower-distribution drug NPs~\cite{zhang2009nanonization, matteucci2008flocculated}. Thermal control also governs mesoscale structure: Higuchi et al.~\cite{higuchi2010phase} showed that thermal annealing drives disorder–order and order–order transitions in block-copolymer NPs formed by solvent evaporation, producing lamellar and onion-like morphologies depending on processing temperature (FIG.\ref{FIG5}(b-1)). Rapid thermal treatments such as microwave annealing can reconfigure morphology on minute timescales, demonstrating kinetic pathways distinct from slow thermal equilibration~\cite{higuchi2013reorientation}. Conversely, heating the antisolvent can accelerate solvent diffusion and lower viscosity, facilitating uniform assembly; Wang et al.~\cite{wang2020lipidation} reported that warmed aqueous baths produced spherical, homogeneous lipid–polymer hybrid NPs with high encapsulation efficiency. Recent process-scale examples include evaporation-triggered spinning-disc formation of PLGA NPs at 40$\sim$50$^\circ$C, yielding concentrated dispersions with narrow size suitable for drug delivery~\cite{zander2025evaporation}. To address solvent-residue concerns, all-aqueous thermal strategies~\cite{guerassimoff2024thermosensitive} exploit LCST behavior of thermoresponsive prodrugs: cold-soluble copolymer prodrugs precipitate into stable NPs upon injection into hot water, eliminating organic solvents while avoiding burst release and enabling sustained, combination drug delivery (FIG.\ref{FIG5}(b-2)). 

The advantages of temperature-assisted nanoprecipitation include: enabling new NP morphologies (e.g. vesicles), improving mixing and encapsulation, enhancing solvent removal, and retaining NP’s inherent simplicity. It is often simple and easily implemented, resulting in its current main use in batch nanoprecipitation mixing. It usually yields NPs in a single step with minimal post-treatment. Continued research is expanding thermal-assisted mode to new polymers (like poloxamers, polypeptides) and mixing technologies (microfluidic devices at controlled temperature). Despite its benefits, a key limitation is particle polydispersity and size increase under some conditions for temperature-assisted nanoprecipitation.  Too rapid nucleation can lead to uncontrolled aggregation or incomplete stabilization of nascent NPs. Other disadvantages include need for precise temperature/solvent control, potential thermal degradation of payload, and additional process complexity (venting solvent vapors, scaling thermal baths).    

\subsection{Chemical-assisted nanoprecipitation}

Chemical-assisted nanoprecipitation utilizes simple chemical stimuli (i.e., pH, salt/ionic strength, counterions and small reactive additives) to modulate polymer solubility, intermolecular interactions and phase behavior, thereby directing NP formation without special equipment. These chemical triggers operate by either changing effective solvent quality or by modifying solvent–solute interactions. As a result, chemical inputs provide a versatile, readily tunable route to supersaturation, liquid–liquid demixing and ultimate NP solidification, enabling morphology and encapsulation performances that are difficult to obtain by solvent-shift alone.

pH-triggered nanoprecipitation is founded on polymers or solutes bearing ionizable moieties. Modulation of solution pH alters the degree of protonation of such groups and thus the polymer’s charge, hydrophilicity and colloidal stability~\cite{alvarez2025preparation}. As the pH traverses a threshold, polymers can undergo LCST-like coil–to–globule transitions or abrupt solubility loss, producing a “pH-switch” route to precipitation and self-assembly~\cite{guerassimoff2024thermosensitive}. This mechanism is both mild and, in many systems, reversible, making it well suited to stimuli-responsive carriers: for example, pH-sensitive copolymers (tertiary amines in PDPA or carboxylates in PAA) can be triggered to form vesicular nanostructures at a defined pH and to disassemble upon exposure to acidic microenvironments for cargo release~\cite{stone2025fabrication, pal2022bioactive}. Practical demonstrations include PLGA/carboxymethyl chitosan core–shell NPs that exploit pH responsiveness to differentially load doxorubicin in the shell and docetaxel in the core, producing sequential release profiles~\cite{shanavas2019polymeric}.

\begin{figure*}[!t]
\centering
\includegraphics[width = 0.97\textwidth]{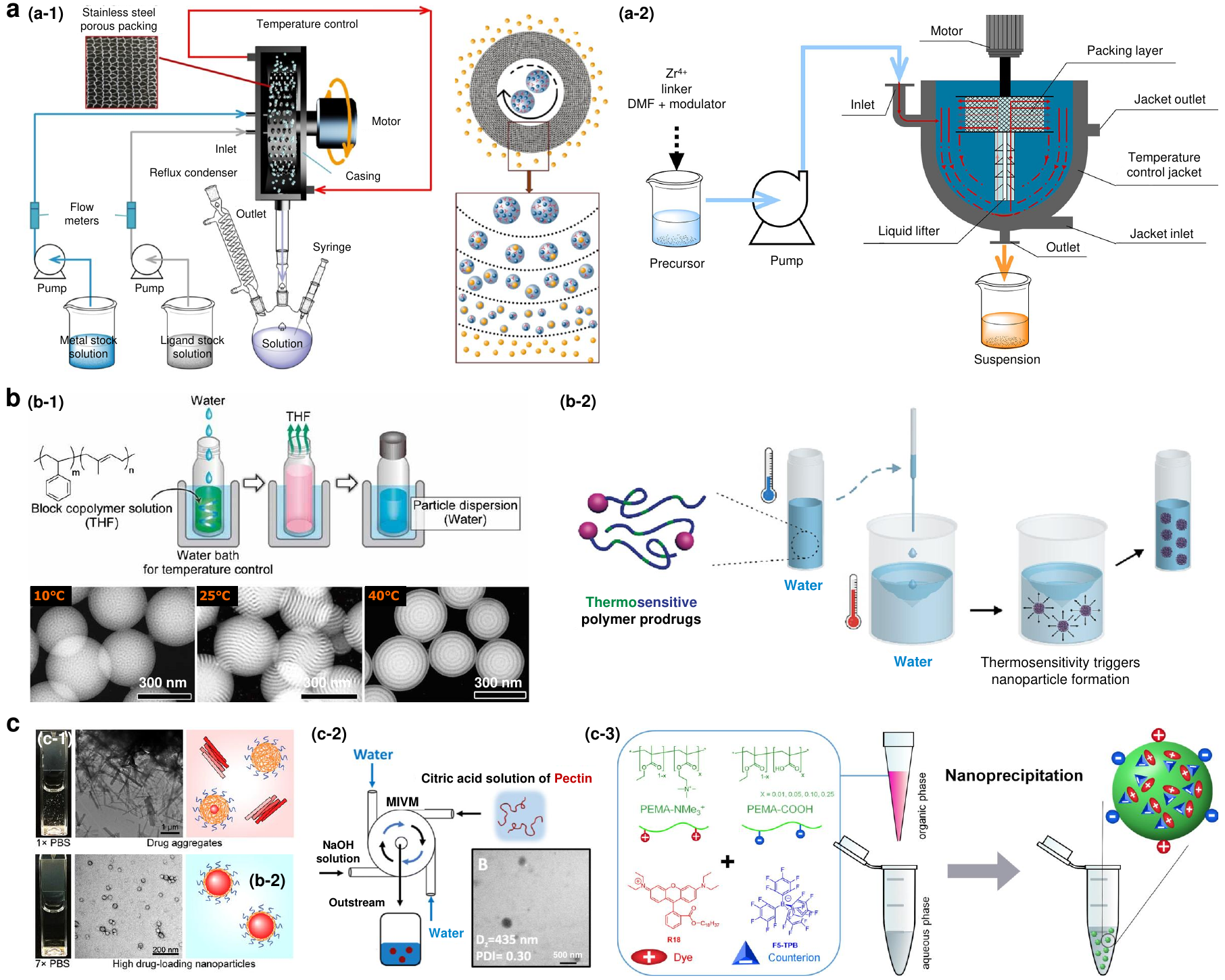}
\caption{(a) Supergravity-assisted nanoprecipitation using RPB. (a-1) synthesis device and schematic of the formation of ultrasmall MOFs in an ECRPB reactor. Adapted from Ref.~\cite{chang2021general}. (a-2) Schematic diagram of the synthetic route in an ICRPB reactor. Adapted from Ref.~\cite{guo2023universal}.  
(b) Thermal-assisted nanoprecipitation with two cases. (b-1) Schematic illustration of the preparation method of block copolymer NPs, and the STEM images of PSt-PI-30 NPs prepared at different temperatures; (b-2) Schematic of the all-aqueous nanoprecipitation process for formation of thermosensitive polymer prodrug NPs. Adapted from Refs.~\cite{higuchi2010phase, guerassimoff2024thermosensitive}. 
(c) Chemical-assisted nanoprecipitation with three cases. (c-1) The snapshot, TEM image, and schematic for the NPs produced by bad salt concentration (top) and good salt concentration (down); (c-2) Schematic illustration of the preparation of pectin NPs through continuous aqueous nanoprecipitation and the TEM image of pectin NPs; (c-3) Preparation of polymer NPs through nanoprecipitation using combinations of polymers bearing oppositely charged groups. Adapted from Refs.~\cite{yang2023phase, ding2023aqueous, combes2022protein}.  
}
\label{FIG5}
\end{figure*}

Salt-assisted nanoprecipitation exploits ionic strength to depress polymer solubility by reducing water activity and perturbing polymer–water interactions. Strongly hydrated ions sequester “free” solvent, frequently inducing liquid–liquid demixing that precedes solid precipitation and thereby permitting controlled delay of polymer collapse to promote drug co-precipitation and higher payloads. Yang et al.~\cite{yang2023phase} demonstrated a salt-screening strategy in which tailored ionic concentrations induced phase separation and substantially enhanced drug loading (FIG.~\ref{FIG5}(c-1)). In practice, salting-out is particularly useful for water-soluble polymers or as a concentration step, but it often yields larger NPs with broader size distributions than rapid solvent-shift methods and may require downstream desalting or crosslinking for colloidal stabilization~\cite{eltaib2025polymeric}. Salt species can also act chemically and play a dual role : for example, HS$^-$ from NaHS both chemically denitrates nitrocellulose and, via ionic strength, promotes salting-out, lowering solubility and triggering nucleation~\cite{yin2025novel}.

Reactive and ion-pairing strategies further broaden the chemical toolkit. Hydrophobic ion pairing uses amphiphilic counterions to render charged solutes temporarily insoluble~\cite{ristroph2019hydrophobic}. A polar drug or polymer carrying ionic groups is complexed with a hydrophobic ion to form a neutral, water-insoluble salt. These ion-paired complexes behave like hydrophobic molecules: they phase-separate rapidly during nanoprecipitation. Lu et al.\cite{lu2018encapsulation} demonstrated efficient encapsulation of peptide:fatty acid complexes during millisecond FNP, while tunability in counterion selection enables encapsulation of proteins and modulation of release profiles\cite{ristroph2021highly}. Reactive FNP has also enabled stabilization of inorganic nanocrystals, such as lanthanum phosphate co-doped with Tb$^{3+}$ and Yb$^{3+}$\cite{fan2021continuous}. In inverse FNP, ionic additives preserve protein stability and activity, as shown by Ca$^{2+}$ maintaining DNase-1 structure and spermine crosslinking PAA, while triethylamine balanced internal pH~\cite{maiocchi2025development}. 

Ionic additives tune surface chemistry and colloidal behavior. For zein NPs, phosphate ions ($\rm Ca(H_2PO_4)_2\cdot H_2O$) altered size, zeta potential, and dispersity; phosphate concentration improved stability, while pH dictated release kinetics~\cite{alvarez2025preparation}. Conversely, acidic conditions promoted aggregation, whereas alkaline environments enhanced hydrophilicity; certain ions (K$^+$, Br$^-$) induced aggregation, while surfactants stabilized colloids~\cite{condello2025two}. Coordination strategies add further control: Fe$^{3+}$–luteolin complexes yield stable, high-loading NPs during FNP process with pH-responsive release and photothermal functionality~\cite{ruan2024polymeric}. Collectively, these approaches demonstrate how chemical triggers can direct nucleation pathways, structural organization, and functional performance of NPs.

Chemical triggers also enable solvent-free or aqueous-only routes. For instance, citric acid/NaOH solvent-shifts can alter pectin solubility and trigger scalable, antisolvent-free formation of biopolymer NPs in aqueous-only nanoprecipitation systems~\cite{ding2023aqueous} (FIG.\ref{FIG5}(c-2)). Mixed-charge polymer NPs, formed by kinetically trapping oppositely charged polymers to prevent full charge pairing, yield sub-25 nm NPs with tunable surface charge, reversible charge switching and efficient encapsulation of hydrophobic dyes, thereby imparting programmable cell-interaction behaviors that mimic protein-like surfaces~\cite{combes2022protein} (FIG.\ref{FIG5}(c-3)). Despite these advantages, chemical-assisted mode is constrained by chemical specificity and tradeoffs in size control: only polymers bearing appropriate responsive groups respond to pH triggers; salts often broaden size distributions and require purification; and ionic additives can induce aggregation unless counterbalanced by steric stabilizers. Nonetheless, by applying straightforward, scalable chemical inputs, these methods offer powerful routes to functional, stimuli-responsive NPs for drug delivery, imaging and advanced materials.

\subsection{Other stimulus-assisted methods and multi-stimulus coupling}

Nanoprecipitation also enables other stimulus-assisted routes for NP synthesis. Several studies use light/photo during or immediately after precipitation to crosslink shells, stabilize vesicles, or convert photo-prodrugs into active payloads, making this strategy particularly valuable for theranostics where illumination both fixes particle structure and activates cargo~\cite{shamsipur2023phototriggered}. Cheng et al.~\cite{cheng2024template} demonstrated a template-free, one-pot continuous gradient nanoprecipitation of incompatible polymers, where UV-assisted interfacial crosslinking yields hollow and multivoid NPs after simple solvent extraction of uncrosslinked material.

As nanomaterials grow more complex, multi-stimulus-assisted nanoprecipitation, combining two or more external triggers during or immediately around precipitation to govern nucleation, growth, morphology, loading, and function, shows great promise. Multi-stimulus coupling effectively builds segmented or hybrid reactors in which different zones or fields act sequentially or simultaneously; coupling can be parallel (dual fields on mixed streams) or serial (one field followed by another). Mechanistic insight is improving but remains partly empirical. For example, FNP paired with immediate phototriggered crosslinking and pH-responsive polymer choice produces carriers that form and are light-stabilized in one step for controlled release and imaging~\cite{kang2023photo}. Patel et al.~\cite{patel2024enhancing} used synchronized temperature control and ultrasound in microfluidic antisolvent nanoprecipitation to produce febuxostat NPs with Poloxamer 407, achieving ultra-small NPs with narrow dispersity and high drug loading~\cite{behera20243d}. Preclinical studies report improved bioavailability and immunogenicity for small molecules and biologics under combined stimuli strategies~\cite{maiocchi2025development, song2023near}.

Future designs will integrate microreactors, high-shear mixers, ultrasound or microwaves into a single "segmented multi-field intensified reaction system"~\cite{feng2025microchannels}. This highlights the direction of combining triggers for next-generation nanoprecipitation. However, a fundamental understanding of multistimulus parameter coupling remains incomplete, i.e., the spatiotemporal interplay between cavitation events, temperature gradients and sub-millisecond mixing, and scaling up is also an engineering challenge.

\newcolumntype{L}{>{\RaggedRight\hangafter=1\hangindent=0em}X}	

	\begin{table*}[htbp]
		\tiny
		\centering
		\renewcommand\arraystretch{2} 
		\label{tab:table1}
		\renewcommand{\tablename}{Table}
		\caption{Comparison of six types of stimulus-assisted nanoprecipitation methods.}
		\setlength{\tabcolsep} {3pt}
		
		\begin{tabularx}{\textwidth}{lp{4.5cm}p{4.3cm}p{6cm} }
				
		\toprule[1pt]
		\fontsize{7}{7}\selectfont \textbf{Stimuli} & \fontsize{7}{7}\selectfont \textbf{Advantages} & \fontsize{7}{7}\selectfont \textbf{Limitations} & \fontsize{7}{7}\selectfont \textbf{Representative applications} \\
		\midrule[0.5pt]

\fontsize{6}{7}\selectfont \textbf{Ultrasonic-} &  \fontsize{6}{7}\selectfont \makecell[tl]{Rapid mixing and nucleation;\\Non-contact and tunable;\\Post-synthesis dispersion;\\Non-mechanical energy and easy to implement;} &  \fontsize{6}{7}\selectfont \makecell[tl]{Potential thermal/chemical stress;\\Energy‐intensive for large volumes;\\Potential degradation;\\Inhomogeneity and scale-up complexity;}  & \fontsize{6}{7}\selectfont \makecell[tl]{Probe-type sonication~\cite{zhang2024physicochemical};\\Bath-type sonication~\cite{wang2022formation};\\GHz-frequency BAW resonators~\cite{xu2024microfluidic};\\Multi-inlet acoustofluidic mixer~\cite{wu2025acoustofluidic};} \\

 \fontsize{6}{7}\selectfont \textbf{Electrical-} &  \fontsize{6}{7}\selectfont \makecell[tl]{High monodispersity with precise size control;\\ Low flow, continuous operation (scalability);\\Enhanced encapsulation efficiency and versatility;\\Faster, more uniform mixing at low $Re$; } &  \fontsize{6}{7}\selectfont \makecell[tl]{Specialized and complex equipment;\\Throughput limits;\\Material and conductivity constraints;\\Post-processing and optimization needed;}  & \fontsize{6}{7}\selectfont \makecell[tl]{Continuous electrohydrodynamic mixing~\cite{lee2024semibatch};\\AC electroosmosis micromixer~\cite{modarres2020electrohydrodynamic};\\Lab-on-a-foil electric microfluidic device~\cite{wu2022ac};\\AC electrothermal wavy micromixer~\cite{mehta2023ac};} \\

 \fontsize{6}{7}\selectfont	\textbf{Supergravity-}    &   \fontsize{6}{7}\selectfont \makecell[tl]{Extreme mixing intensity;\\Small NP sizes and monodispersity;\\ Continuous, scalable and high throughput;\\Reduced fouling without complex chemistry;} &  \fontsize{6}{7}\selectfont \makecell[tl]{Specialized equipment and cost;\\ More costly than simple mixers;\\Limited viscosity and material constraints;\\Design and parameter sensitivity;} & \fontsize{6}{7}\selectfont \makecell[tl]{Rotating packed bed for MOFs~\cite{chang2021general};\\High gravity anti-solvent precipitation~\cite{dong2024instantaneous};\\Internal circulation RPB~\cite{guo2023universal};\\High-gravity reactive precipitation~\cite{fang2021preparation};}  \\

 \fontsize{6}{7}\selectfont \textbf{Thermal-} & \fontsize{6}{7}\selectfont \makecell[tl]{Expanded material versatility;\\Simple, precise and easily implemented;\\Higher uniformity and encapsulation;\\Potential for sequential control;}   & \fontsize{6}{7}\selectfont  \makecell[tl]{Increased NP heterogeneity;\\Mild effect on nucleation;\\Potential thermal degradation or stress;\\Scale-up and reproducibility;}  & \fontsize{6}{7}\selectfont \makecell[tl]{Phase transformation of block-copolymer NPs~\cite{higuchi2010phase};\\Lipid–polymer hybrid NPs~\cite{wang2020lipidation};\\Poly(ethylene oxide)-block-polycaprolactone NPs~\cite{kankanen2020evaluation};\\ Thermosensitive polymer prodrug NPs~\cite{guerassimoff2024thermosensitive};} \\
        
\fontsize{6}{7}\selectfont \textbf{Chemical-} & \fontsize{6}{7}\selectfont \makecell[tl]{Mild, green conditions;\\Selective or reversible triggering;\\ High drug loading;\\ Easy to implement in batch or continuous flow;}    & \fontsize{6}{7}\selectfont \makecell[tl]{Chemical specificity and narrow applicability;\\Control challenges;\\ Broader polydispersity;\\Residual reagents and possible aggregation;}  & \fontsize{6}{7}\selectfont \makecell[tl]{Salt-screening strategy~\cite{yang2023phase};\\Hydrophobic ion pairing~\cite{ristroph2021highly};\\ Aqueous-only nanoprecipitation~\cite{ding2023aqueous};\\ Mixed-charge polymer NPs~\cite{combes2022protein};}   \\
		
 \fontsize{6}{7}\selectfont	\textbf{Multistimulus-} &  \fontsize{6}{7}\selectfont \makecell[tl]{Extended control over nucleation and morphology;\\ Enhanced encapsulation and functionalization;\\ Integration with continuous production; \\ Faster processing and smaller sizes; } & \fontsize{6}{7}\selectfont \makecell[tl]{High complexity;\\ Payload and material sensitivity;\\Reproducibility and scale-up;\\ Optimization difficulty; } & \fontsize{6}{7}\selectfont \makecell[tl]{Phototriggered crosslinking and pH-responsive polymer~\cite{kang2023photo};\\Temperature$\&$ultrasound in microfluidic platform~\cite{patel2024enhancing};\\ \\ } \\
		\bottomrule[1pt]
	\end{tabularx}
	\end{table*}%

\section{Conclusions and outlook}

Nanoprecipitation is the most widely employed bottom-up NP synthesis method due to its rapidity, simplicity, and reproducibility. Stimulus-assisted nanoprecipitation, however, represents a frontier in nanomaterial synthesis, where external triggers beyond simple mixing are exploited to tailor NPs. In this review, we mainly covered six categories of such external triggers: ultrasonic-, electrical-, supergravity-, thermal-, chemical-, and others (stimuli for specific use and multi-stimulus coupling). The comparison of these six methods is shown in Table~\ref{tab:table1}. The coupling of various trigger methods and mixing approaches/devices faces numerous mechanistic challenges. 
Generally, ultrasonic stimulation is effectively coupled with batch, microfluidic, and FNP systems, leveraging cavitation and acoustic streaming to enhance mixing and nucleation. Electrical assistance is primarily implemented in microfluidic platforms via electrohydrodynamic mixing or embedded electrodes, as well as in specialized electrospray setups, enabling precise control at low flow rates. Supergravity stimulation is uniquely and effectively applied in RPB reactors, which generate intense mixing under high-gravity fields for continuous, nucleation-dominated synthesis. Thermal stimulation is a widely adaptable trigger, commonly used in simple batch processes and increasingly integrated into controlled microfluidic and FNP systems to regulate solvent dynamics and polymer chain mobility. Chemical stimulation, through pH, ionic strength, or additives, is a flexible approach compatible with batch mixing, FNP, and microfluidic methods, often enabling solvent-free or aqueous-only routes. Finally, other stimuli like light and, more promimently, multi-stimulus coupling are advanced strategies that combine forces, such as ultrasound with temperature in microfluidics or phototriggering with FNP, within segmented or hybrid reactor systems to achieve superior control over NP structure and function, representing the frontier of stimulus-assisted nanoprecipitation. However, flash and microfluidic nanoprecipitation processes induce extremely rapid and precisely time adjustable mixing and may reduce the sensitivity of the stimulus to exert its effect or narrow the window of effective action. Further developments exploiting these continuous processes offer a horizon full of promise. How to avoid additional damage caused by these stimuli is also a hot topic for further research. 

While structural and functional optimizations have been achieved, future research should continue to explore the mechanistic synergies of triggers, strive for greener and scalable processes, and target novel applications (e.g. multi-functional drug carriers, precision catalysts, or smart sensors) that leverage the unique control offered by these methods. NPs with structural or chemical anisotropy are promising materials in domains as diverse as cellular delivery, photonic materials, or interfacial engineering. Introducing reactivity into such polymeric nanomaterials is thus of great potential, which brings new challenges to nanoprecipitation technology. 

Besides, the explosive development of machine learning and even artificial intelligence (AI) in recent years has opened up new routes for the use of nanoprecipitation technology in NP synthesis, which can be called: machine learning-assisted nanoprecipitation or AI-assisted nanoprecipitation. AI has been applied towards numerous areas of nanomaterial synthesis~\cite{di2023machine, kuddushi2025recent}. This field is emerging and provides an opportunity for researchers to leverage AI to improve nanoprecipitation techniques.


\section*{Declaration of Competing Interest}

The authors declare that they have no known competing financial interests or personal relationships that could have appeared to influence the work reported in this paper.

\section*{Acknowledgments}

This work is supported by National Natural Science Foundation of China under Grants Nos. 12572290 and 12202244, and the Oceanic Interdisciplinary Program of Shanghai Jiao Tong University (No. SL2023MS002).

\section*{Data availability}

The authors do not have permission to share data.



\bibliographystyle{elsarticle-num}




\section*{References and recommended reading}

Papers of particular interest, published within the period of review, have been highlighted as:\\
*  of special interest \\
** of outstanding interest


\begin{thebibliography}{10}
\expandafter\ifx\csname url\endcsname\relax
  \def\url#1{\texttt{#1}}\fi
\expandafter\ifx\csname urlprefix\endcsname\relax\def\urlprefix{URL }\fi
\expandafter\ifx\csname href\endcsname\relax
  \def\href#1#2{#2} \def\path#1{#1}\fi






\bibitem{yan2021nanoprecipitation}
X.~Yan, J.~Bernard, F.~Ganachaud, Nanoprecipitation as a simple and
  straightforward process to create complex polymeric colloidal morphologies,
  Advances in Colloid and Interface Science 294 (2021) 102474.

\bibitem{kuddushi2025recent}
M.~Kuddushi, C.~Kanike, B.~B. Xu, X.~Zhang, Recent advances in
  nanoprecipitation: from mechanistic insights to applications in nanomaterial
  synthesis, Soft Matter 21~(15) (2025) 2759--2781.\\
\textcolor{blue}{* This review paper systematically summarizes recent advances in nanoprecipitation from mechanistic insights to applications in nanomaterial synthesis.}

\bibitem{fesenmeier2024overcoming}
D.~J. Fesenmeier, E.~S. Cooper, Y.-Y. Won, Overcoming limitations of
  conventional solvent exchange methods: Achieving monodisperse non-equilibrium
  polymer micelles through equilibration-nanoprecipitation (enp), Journal of
  Colloid and Interface Science 661 (2024) 861--869.

\bibitem{liu2020stable}
Y.~Liu, G.~Yang, T.~Baby, Tengjisi, D.~Chen, D.~A. Weitz, C.-X. Zhao, Stable
  polymer nanoparticles with exceptionally high drug loading by sequential
  nanoprecipitation, Angewandte Chemie 132~(12) (2020) 4750--4758.

\bibitem{yan2019programmable}
X.~Yan, R.~A. N.~S. Ramos, P.~Alcouffe, L.~E. Munoz, R.~O. Bilyy, F.~Ganachaud,
  J.~Bernard, Programmable hierarchical construction of mixed/multilayered
  polysaccharide nanocapsules through simultaneous/sequential nanoprecipitation
  steps, Biomacromolecules 20~(10) (2019) 3915--3923.

\bibitem{li2025multistage}
S.~Li, B.~Yang, L.~Ye, S.~Hu, B.~Li, Y.~Yang, Y.~Hu, X.~Jia, L.~Feng, Z.~Xiong,
  Multistage microfluidic assisted co-delivery platform for dual-agent facile
  sequential encapsulation, European Journal of Pharmaceutics and
  Biopharmaceutics 207 (2025) 114616. \\
\textcolor{blue}{** This paper reports a multi-stage microfluidic nanoprecipitation platform, referred to as the $TrH$ chip, designed to efficiently and sequentially encapsulate two drugs with distinct properties within a microchannel by precisely controlling the multiphase flow mixing process. } 

\bibitem{fu2020direct}
Z.~Fu, L.~Li, Y.~Wang, Q.~Chen, F.~Zhao, L.~Dai, Z.~Chen, D.~Liu, X.~Guo,
  Direct preparation of drug-loaded mesoporous silica nanoparticles by
  sequential flash nanoprecipitation, Chemical Engineering Journal 382 (2020)
  122905.

\bibitem{lewis2025process}
P.~K. Lewis, N.~El~Amri, E.~E. Burnham, N.~Arrus, N.~M. Pinkerton, Process and
  formulation parameters governing polymeric microparticle formation via
  sequential nanoprecipitation (snap), ACS Engineering Au (2025).

\bibitem{chen2023recent}
T.~Chen, Y.~Peng, M.~Qiu, C.~Yi, Z.~Xu, Recent advances in mixing-induced
  nanoprecipitation: from creating complex nanostructures to emerging
  applications beyond biomedicine, Nanoscale 15~(8) (2023) 3594--3609.

\bibitem{sun2021diverse}
Z.~Sun, B.~Wu, Y.~Ren, Z.~Wang, C.-X. Zhao, M.~Hai, D.~A. Weitz, D.~Chen,
  Diverse particle carriers prepared by co-precipitation and phase separation:
  Formation and applications, ChemPlusChem 86~(1) (2021) 49--58.

\bibitem{sharratt2021precision}
W.~N. Sharratt, V.~E. Lee, R.~D. Priestley, J.~T. Cabral, Precision polymer
  particles by flash nanoprecipitation and microfluidic droplet extraction, ACS
  Applied Polymer Materials 3~(10) (2021) 4746--4768.

\bibitem{bovone2022solvent}
G.~Bovone, L.~Cousin, F.~Steiner, M.~W. Tibbitt, Solvent controls nanoparticle
  size during nanoprecipitation by limiting block copolymer assembly,
  Macromolecules 55~(18) (2022) 8040--8048.

\bibitem{misra2024flash}
B.~Misra, K.~A. Hughes, W.~H. Pentz, P.~Samart, W.~J. Geldenhuys, S.~Bobbala,
  Flash nanoprecipitation assisted self-assembly of ionizable lipid
  nanoparticles for nucleic acid delivery, Nanoscale 16~(14) (2024) 6939--6948.

\bibitem{qi2025synthesis}
L.~Qi, D.~Huang, H.~Kou, A.~Chernatynskaya, N.~Ercal, H.~Yang, Synthesis and
  characterization of free radical scavenging dendrimer nanogels via
  cross-linking reaction-enabled flash nanoprecipitation, Biomacromolecules
  26~(5) (2025) 2986--2995.

\bibitem{manohar2024drying}
A.~Manohar, S.~M, M.~G. Basavaraj, S.~Sudhakar, E.~Mani, Drying-induced flash
  nanoprecipitation in a sessile drop: A route to synthesize polymeric
  nanoparticles, Langmuir 40~(26) (2024) 13613--13621.

\bibitem{vandenberg2025learning}
M.~A. VandenBerg, X.~Dong, W.~C. Smith, G.~Tian, O.~Stephens, T.~F. O’Connor,
  X.~Xu, Learning from the future: towards continuous manufacturing of
  nanomaterials, AAPS Open 11~(1) (2025) 7.

\bibitem{piacentini2022membrane}
E.~Piacentini, B.~Russo, F.~Bazzarelli, L.~Giorno, Membrane nanoprecipitation:
  From basics to technology development, Journal of Membrane Science 654 (2022)
  120564.

\bibitem{mens2024spinning}
W.~Mens, Spinning disc reactor: Status, challenges and future perspectives for
  advancement, Chemical Engineering and Processing-Process Intensification 205
  (2024) 109977.

\bibitem{zander2025evaporation}
A.~J. Zander, M.-S. Ehrlich, S.~ur~Rehman, M.~Schneider, Evaporation-triggered
  nanoprecipitation for plga nanoparticle formation using a spinning-disc
  system, Journal of Drug Delivery Science and Technology 108 (2025) 106901.

\bibitem{tang2025research}
N.~Tang, J.~Wu, J.~Guo, Y.~Wang, H.~Li, J.~Wu, J.~Zhang, Research on continuous
  synthesis of nanoparticles using an inline high-shear mixer: Barium sulfate,
  Industrial \& Engineering Chemistry Research 64~(13) (2025) 7189--7199.

\bibitem{feng2025microchannels}
Z.~Feng, J.~Guo, Y.~Wang, J.~Shi, H.~Shi, H.~Li, J.~Zhang, J.~Wu, From
  microchannels to high shear reactors: process intensification strategies for
  controlled nanomaterial synthesis, Nanoscale Horizons (2025).

\bibitem{sharma2019effect}
C.~Sharma, M.~A. Desai, S.~R. Patel, Effect of surfactants and polymers on
  morphology and particle size of telmisartan in ultrasound-assisted
  anti-solvent crystallization, Chemical Papers 73~(7) (2019) 1685--1694.

\bibitem{benhabiles2025performance}
N.~Benhabiles, N.~Boudries, H.~Mokrane, F.~Chaput, The performance of combined
  ultrasound and nonsolvent precipitation for betalain encapsulation in
  different types of starches: thermal stability and antioxidant activity,
  Journal of Food Measurement and Characterization 19~(1) (2025) 276--292.

\bibitem{huang2019acoustofluidic}
P.-H. Huang, S.~Zhao, H.~Bachman, N.~Nama, Z.~Li, C.~Chen, S.~Yang, M.~Wu,
  S.~P. Zhang, T.~J. Huang, Acoustofluidic synthesis of particulate
  nanomaterials, Advanced Science 6~(19) (2019) 1900913.

\bibitem{zhang2024physicochemical}
X.~Zhang, Q.~Li, J.~Wang, J.~Huang, W.~Huang, Y.~Huang, Physicochemical
  properties and in vitro release of formononetin nano-particles by ultrasonic
  probe-assisted precipitation in four polar organic solvents, Food Chemistry
  461 (2024) 140918.

\bibitem{wang2022formation}
J.~Wang, Y.-D. Yu, Z.-G. Zhang, W.-C. Wu, P.-L. Sun, M.~Cai, K.~Yang, Formation
  of sweet potato starch nanoparticles by ultrasonic—assisted
  nanoprecipitation: Effect of cold plasma treatment, Frontiers in
  Bioengineering and Biotechnology 10 (2022) 986033.

\bibitem{zhang2024antisolvent}
X.~Zhang, Y.~Huang, S.~Huang, W.~Xie, W.~Huang, Y.~Chen, Q.~Li, F.~Zeng,
  X.~Liu, Antisolvent precipitation for the synergistic preparation of
  ultrafine particles of nobiletin under ultrasonication-homogenization and
  evaluation of the inhibitory effects of $\alpha$-glucosidase and porcine
  pancreatic lipase in vitro, Ultrasonics Sonochemistry 105 (2024) 106865.

\bibitem{kamat2023active}
V.~Kamat, P.~Dey, D.~Bodas, A.~Kaushik, A.~Boymelgreen, S.~Bhansali, Active
  microfluidic reactor-assisted controlled synthesis of nanoparticles and
  related potential biomedical applications, Journal of Materials Chemistry B
  11~(25) (2023) 5650--5667.

\bibitem{chen2021sharp}
Z.~Chen, P.~Liu, X.~Zhao, L.~Huang, Y.~Xiao, Y.~Zhang, J.~Zhang, N.~Hao,
  Sharp-edge acoustic microfluidics: Principles, structures, and applications,
  Applied Materials Today 25 (2021) 101239.

\bibitem{ozcelik2014acoustofluidic}
A.~Ozcelik, D.~Ahmed, Y.~Xie, N.~Nama, Z.~Qu, A.~A. Nawaz, T.~J. Huang, An
  acoustofluidic micromixer via bubble inception and cavitation from
  microchannel sidewalls, Analytical chemistry 86~(10) (2014) 5083--5088.

\bibitem{rasouli2019ultra}
M.~R. Rasouli, M.~Tabrizian, An ultra-rapid acoustic micromixer for synthesis
  of organic nanoparticles, Lab on a Chip 19~(19) (2019) 3316--3325.

\bibitem{agha2024integration}
A.~Agha, E.~Abu-Nada, A.~Alazzam, Integration of acoustic micromixing with
  cyclic olefin copolymer microfluidics for enhanced lab-on-a-chip applications
  in nanoscale liposome synthesis, Biofabrication 16~(4) (2024) 045004.

\bibitem{bachman2020acoustofluidic}
H.~Bachman, C.~Chen, J.~Rufo, S.~Zhao, S.~Yang, Z.~Tian, N.~Nama, P.-H. Huang,
  T.~J. Huang, An acoustofluidic device for efficient mixing over a wide range
  of flow rates, Lab on a Chip 20~(7) (2020) 1238--1248.

\bibitem{xu2024microfluidic}
H.~Xu, Z.~Wang, W.~Wei, T.~Li, X.~Duan, Microfluidic confined acoustic
  streaming vortex for liposome synthesis, Lab on a Chip 24~(10) (2024)
  2802--2810. \\
\textcolor{blue}{** This paper reports a GHz-frequency BAW resonator that exploits high-frequency energy confinement and rapid dissipation to drive body-force-driven microvortices without complex microfeatures, thereby reducing clogging risk and enabling flow-rate–independent tuning of liposome size through acoustic power modulation alone.}  

\bibitem{lu2024vortex}
Y.~Lu, W.~Tan, S.~Mu, G.~Zhu, Vortex-enhanced microfluidic chip for efficient
  mixing and particle capturing combining acoustics with inertia, Analytical
  Chemistry 96~(9) (2024) 3859--3869. \\
\textcolor{blue}{* This paper developed an acousto-inertial microfluidic chip that integrates bubble-sharp-edge composite microstructures and contraction–expansion arrays to synergistically enhance vortex generation, enabling efficient fluid mixing and particle manipulation across a wide flow range at low excitation voltages.  }

\bibitem{wu2025acoustofluidic}
X.~Wu, B.~Jin, Y.~Zeng, M.~Wen, Y.~Zou, L.~Lyu, X.~Li, S.~Yan, K.~Zeng,
  M.~Yang, et~al., Acoustofluidic fabrication of calcium pyrophosphate-enzyme
  nanocatalysts for oral treatment of metabolic disorders, ACS nano 19~(28)
  (2025) 25890--25903.

\bibitem{rasouli2023acoustofluidics}
R.~Rasouli, K.~M. Villegas, M.~Tabrizian, Acoustofluidics--changing paradigm in
  tissue engineering, therapeutics development, and biosensing, Lab on a Chip
  23~(5) (2023) 1300--1338.

\bibitem{buyukkanber2024high}
K.~B{\"u}y{\"u}kkanber, A.~Ekinci, {\"O}.~{\c{S}}ahin, High catalytic activity
  of cobalt nanoparticles synthesized by ultrasonic spray method in sodium
  borohydride hydrolysis, International Journal of Hydrogen Energy 79 (2024)
  335--345.

\bibitem{lee2024semibatch}
K.~H. Lee, F.~N. Khan, I.~B. Cosmin, D.~U. Mualen, T.~K. Porter, B.~E.
  Wyslouzil, J.~O. Winter, Semibatch and continuous electrohydrodynamic mixing
  nanoprecipitation for scalable polymer nanostructure production, ACS Applied
  Polymer Materials 6~(20) (2024) 12382--12393.\\
\textcolor{blue}{** This paper reports an electrohydrodynamic mixing nanoprecipitation (EMNP) platform that uses electric voltage instead of high mechanical energy for scalable block copolymer nanoparticle production, enabling continuous operation and size tuning via solvent ratio adjustment without altering polymer properties.}

\bibitem{luo2015preparation}
C.~Luo, T.~Okubo, M.~Nangrejo, M.~Edirisinghe, Preparation of polymeric
  nanoparticles by novel electrospray nanoprecipitation, Polymer International
  64~(2) (2015) 183--187.

\bibitem{rostamabadi2021electrospraying}
H.~Rostamabadi, S.~R. Falsafi, M.~M. Rostamabadi, E.~Assadpour, S.~M. Jafari,
  Electrospraying as a novel process for the synthesis of
  particles/nanoparticles loaded with poorly water-soluble bioactive molecules,
  Advances in colloid and interface science 290 (2021) 102384.

\bibitem{roshan2024curcumin}
Z.~Roshan, V.~Haddadi-Asl, H.~Ahmadi, M.~Moussaei, Curcumin-encapsulated poly
  (lactic-co-glycolic acid) nanoparticles: a comparison of drug release
  kinetics from particles prepared via electrospray and nanoprecipitation,
  Macromolecular Materials and Engineering 309~(7) (2024) 2400040.

\bibitem{zhang2022nano}
C.~Zhang, X.~Wang, M.~Xiao, J.~Ma, Y.~Qu, L.~Zou, J.~Zhang, Nano-in-micro
  alginate/chitosan hydrogel via electrospray technology for orally curcumin
  delivery to effectively alleviate ulcerative colitis, Materials \& Design 221
  (2022) 110894.

\bibitem{sun2016controlled}
X.-T. Sun, C.-G. Yang, Z.-R. Xu, Controlled production of size-tunable janus
  droplets for submicron particle synthesis using an electrospray microfluidic
  chip, RSC advances 6~(15) (2016) 12042--12047.

\bibitem{wang2025electrosprayed}
C.~Wang, S.~G. Gaurkhede, M.~Liu, Y.~Zhou, Electrosprayed micro-to-nanoscale
  particles with tunable morphologies and compositions for pulmonary drug
  delivery applications and inhalation therapies: A review, ACS Biomaterials
  Science \& Engineering (2025).

\bibitem{wu2025design}
M.~Wu, Z.~Liu, Y.~Gao, Design and fabrication of microelectrodes for
  dielectrophoresis and electroosmosis in microsystems for bio-applications,
  Micromachines 16~(2) (2025) 190.

\bibitem{modarres2020electrohydrodynamic}
P.~Modarres, M.~Tabrizian, Electrohydrodynamic-driven micromixing for the
  synthesis of highly monodisperse nanoscale liposomes, ACS Applied Nano
  Materials 3~(5) (2020) 4000--4013.

\bibitem{modarres2020phase}
P.~Modarres, M.~Tabrizian, Phase-controlled field-effect micromixing using ac
  electroosmosis, Microsystems \& Nanoengineering 6~(1) (2020) 60.

\bibitem{wu2022ac}
M.~Wu, Y.~Gao, A.~Ghaznavi, W.~Zhao, J.~Xu, Ac electroosmosis micromixing on a
  lab-on-a-foil electric microfluidic device, Sensors and Actuators B: Chemical
  359 (2022) 131611.

\bibitem{mehta2023ac}
S.~K. Mehta, P.~K. Mondal, Ac electrothermal effect promotes enhanced solute
  mixing in a wavy microchannel, Langmuir 39~(47) (2023) 16797--16806.\\
\textcolor{blue}{* This paper developed an AC electrothermal-driven wavy microchannel mixer that generates enhanced three-dimensional vortices for efficient solute mixing with minimal temperature rise, showing potential for improving nanoprecipitation processes requiring rapid and uniform mixing.        } 

\bibitem{lee2021polymer}
K.~H. Lee, F.~N. Khan, L.~Cosby, G.~Yang, J.~O. Winter, Polymer concentration
  maximizes encapsulation efficiency in electrohydrodynamic mixing
  nanoprecipitation, Frontiers in nanotechnology 3 (2021) 719710.

\bibitem{chatterjee2020electrospray}
M.~Chatterjee, R.~Maity, S.~Das, N.~Mahata, B.~Basu, N.~Chanda,
  Electrospray-based synthesis of fluorescent poly (d, l-lactide-co-glycolide)
  nanoparticles for the efficient delivery of an anticancer drug and
  self-monitoring its effect in drug-resistant breast cancer cells, Materials
  Advances 1~(8) (2020) 3033--3048.

\bibitem{chen2000synthesis}
J.-F. Chen, Y.-H. Wang, F.~Guo, X.-M. Wang, C.~Zheng, Synthesis of
  nanoparticles with novel technology: high-gravity reactive precipitation,
  Industrial \& engineering chemistry research 39~(4) (2000) 948--954.

\bibitem{tao2024highly}
C.~Tao, F.~Li, Z.~Ma, X.~Li, Y.~Zhang, Y.~Le, J.~Wang, J.~Zhao, C.~Liu,
  J.~Zhang, Highly efficient oral iguratimod/polyvinyl alcohol nanodrugs
  fabricated by high-gravity nanoprecipitation technique for treatment of
  rheumatoid arthritis, Small 20~(13) (2024) 2304150.

\bibitem{liu2019cfd}
Y.~Liu, W.~Wu, Y.~Luo, G.-W. Chu, W.~Liu, B.-C. Sun, J.-F. Chen, Cfd simulation
  and high-speed photography of liquid flow in the outer cavity zone of a
  rotating packed bed reactor, Industrial \& Engineering Chemistry Research
  58~(13) (2019) 5280--5290.

\bibitem{chang2021general}
M.~Chang, Y.~Wei, D.~Liu, J.-X. Wang, J.-F. Chen, A general strategy for
  instantaneous and continuous synthesis of ultrasmall metal--organic framework
  nanoparticles, Angewandte Chemie 133~(50) (2021) 26594--26600.

\bibitem{fang2021preparation}
L.~Fang, Q.~Sun, Y.-H. Duan, J.~Zhai, D.~Wang, J.-X. Wang, Preparation of
  transparent baso4 nanodispersions by high-gravity reactive precipitation
  combined with surface modification for transparent x-ray shielding
  nanocomposite films, Frontiers of Chemical Science and Engineering 15~(4)
  (2021) 902--912.

\bibitem{guo2023universal}
Y.~Guo, Y.~Wei, M.~Chang, D.~Wang, J.-X. Wang, J.-F. Chen, A universal strategy
  for efficient synthesis of zr-based mof nanoparticles for enhanced water
  adsorption, AIChE Journal 69~(9) (2023) e18181.\\
\textcolor{blue}{** This paper developed an internal circulation rotating packed bed (ICRPB) reactor for the efficient, tunable, and universal synthesis of Zr-MOF nanoparticles via enhanced nanoscale precipitation under high-gravity conditions.    }

\bibitem{rahmani2015long}
S.~Rahmani, C.~H. Villa, A.~F. Dishman, M.~E. Grabowski, D.~C. Pan, H.~Durmaz,
  A.~C. Misra, L.~Col{\'o}n-Mel{\'e}ndez, M.~J. Solomon, V.~R. Muzykantov,
  et~al., Long-circulating janus nanoparticles made by electrohydrodynamic
  co-jetting for systemic drug delivery applications, Journal of drug targeting
  23~(7-8) (2015) 750--758.

\bibitem{dong2024instantaneous}
Y.-J. Dong, J.~Lu, M.~Qiao, K.~Wang, Y.~Le, J.-X. Wang, J.-F. Chen,
  Instantaneous and continuous preparation of avermectin nanoparticles with
  high gravity anti-solvent precipitation technology, Industrial \& Engineering
  Chemistry Research 63~(4) (2024) 1729--1739.

\bibitem{zhang2024preparation}
Y.~Zhang, X.~Wang, J.~Zhang, Preparation of tofacitinib-loaded poly
  (lactic-co-glycolic acid) sustained release nanoparticles by high-gravity
  nanoprecipitation technique and its performance in rheumatoid arthritis,
  Industrial \& Engineering Chemistry Research 63~(29) (2024) 12766--12777.

\bibitem{wang2022surfactant}
M.~Wang, R.~Fan, J.~Zhang, L.~Li, J.-X. Wang, Y.~Le, Surfactant-free synthesis
  of pnipam-based smart microgels for drug delivery using a high-gravity
  rotating packed bed, Industrial \& Engineering Chemistry Research 61~(49)
  (2022) 17866--17875.

\bibitem{wang2025synthesis}
A.-S. Wang, J.-X. Wang, Y.~Le, D.~Wang, Y.~Pu, X.-F. Zeng, J.-F. Chen,
  Synthesis of monodispersed inorganic nanoparticles by high gravity technology
  for multifunctional applications, Current Opinion in Chemical Engineering 47
  (2025) 101060.\\
\textcolor{blue}{* This review introduces the advancements in the synthesis of monodispersed inorganic nanoparticles (metals, metal oxides, and inorganic salts) by high gravity technology for multifunctional applications. }

\bibitem{li2023thermal}
M.~Li, Y.~Wakata, H.~Zeng, C.~Sun, On the thermal response of multiscale
  nanodomains formed in trans-anethol/ethanol/water surfactant-free
  microemulsion, Journal of Colloid and Interface Science 652 (2023)
  1944--1953.

\bibitem{zhou2017polymeric}
J.~Zhou, R.~Ni, Y.~Chau, Polymeric vesicle formation via temperature-assisted
  nanoprecipitation, RSC Advances 7~(29) (2017) 17997--18000.

\bibitem{kankanen2020evaluation}
V.~K{\"a}nk{\"a}nen, J.~Seitsonen, H.~Tuovinen, J.~Ruokolainen, J.~Hirvonen,
  V.~Balasubramanian, H.~A. Santos, Evaluation of the effects of
  nanoprecipitation process parameters on the size and morphology of poly
  (ethylene oxide)-block-polycaprolactone nanostructures, International Journal
  of Pharmaceutics 590 (2020) 119900.

\bibitem{yan2017effect}
X.~Yan, Y.~Chang, Q.~Wang, Y.~Fu, J.~Zhou, Effect of drying conditions on
  crystallinity of amylose nanoparticles prepared by nanoprecipitation,
  International Journal of Biological Macromolecules 97 (2017) 481--488.

\bibitem{huang2018tuning}
W.~Huang, C.~Zhang, Tuning the size of poly (lactic-co-glycolic acid)(plga)
  nanoparticles fabricated by nanoprecipitation, Biotechnology journal 13~(1)
  (2018) 1700203.

\bibitem{wang2020lipidation}
J.~Wang, H.~Zope, M.~A. Islam, J.~Rice, S.~Dodman, K.~Lipert, Y.~Chen, B.~R.
  Zetter, J.~Shi, Lipidation approaches potentiate adjuvant-pulsed immune
  surveillance: a design rationale for cancer nanovaccine, Frontiers in
  Bioengineering and Biotechnology 8 (2020) 787.

\bibitem{tao2019application}
J.~Tao, S.~F. Chow, Y.~Zheng, Application of flash nanoprecipitation to
  fabricate poorly water-soluble drug nanoparticles, Acta pharmaceutica sinica
  B 9~(1) (2019) 4--18.

\bibitem{perumal2022review}
S.~Perumal, R.~Atchudan, W.~Lee, A review of polymeric micelles and their
  applications, Polymers 14~(12) (2022) 2510.

\bibitem{zhang2009nanonization}
Z.-B. Zhang, Z.-G. Shen, J.-X. Wang, H.~Zhao, J.-F. Chen, J.~Yun, Nanonization
  of megestrol acetate by liquid precipitation, Industrial \& engineering
  chemistry research 48~(18) (2009) 8493--8499.

\bibitem{matteucci2008flocculated}
M.~E. Matteucci, J.~C. Paguio, M.~A. Miller, R.~O. Williams~III, K.~P.
  Johnston, Flocculated amorphous nanoparticles for highly supersaturated
  solutions, Pharmaceutical research 25~(11) (2008) 2477--2487.

\bibitem{higuchi2010phase}
T.~Higuchi, K.~Motoyoshi, H.~Sugimori, H.~Jinnai, H.~Yabu, M.~Shimomura, Phase
  transition and phase transformation in block copolymer nanoparticles,
  Macromolecular rapid communications 31~(20) (2010) 1773--1778.

\bibitem{higuchi2013reorientation}
T.~Higuchi, M.~Shimomura, H.~Yabu, Reorientation of microphase-separated
  structures in water-suspended block copolymer nanoparticles through microwave
  annealing, Macromolecules 46~(10) (2013) 4064--4068.

\bibitem{guerassimoff2024thermosensitive}
L.~Guerassimoff, M.~Ferrere, S.~Van~Herck, S.~Dehissi, V.~Nicolas, B.~G.
  De~Geest, J.~Nicolas, Thermosensitive polymer prodrug nanoparticles prepared
  by an all-aqueous nanoprecipitation process and application to combination
  therapy, Journal of Controlled Release 369 (2024) 376--393.\\
\textcolor{blue}{** This paper reports a thermosensitive polymer prodrug nanoparticle platform that utilizes an all-aqueous nanoprecipitation process triggered by temperature-induced solubility transition, enabling organic solvent-free formulation, sustained drug release, and effective combination therapy. }

\bibitem{alvarez2025preparation}
M.~G. Alvarez-Moreno, F.~Rodr{\'\i}guez-F{\'e}lix, C.~G. Barreras-Urbina,
  M.~Plascencia-Jatomea, E.~O. Rueda-Puente, J.~J. Reyes-P{\'e}rez, J.~A.
  Tapia-Hern{\'a}ndez, S.~E. Burruel-Ibarra, T.~J. Madera-Santana, I.~Y.
  L{\'o}pez-Pe{\~n}a, et~al., Preparation and characterization of
  zein-phosphate nanoparticles by nanoprecipitation method with potential use
  as fertilizer, ACS omega (2025).

\bibitem{stone2025fabrication}
P.~T. Stone, A.~J. Kwiatkowski, E.~W. Roth, O.~Fedorova, A.~M. Pyle, J.~T.
  Wilson, Fabrication of rig-i-activating nanoparticles for intratumoral
  immunotherapy via flash nanoprecipitation, Molecular Pharmaceutics (2025).

\bibitem{pal2022bioactive}
V.~K. Pal, S.~Roy, Bioactive peptide nano-assemblies with ph-triggered shape
  transformation for antibacterial therapy, ACS Applied Nano Materials 5~(8)
  (2022) 12019--12034.

\bibitem{shanavas2019polymeric}
A.~Shanavas, N.~K. Jain, N.~Kaur, D.~Thummuri, M.~Prasanna, R.~Prasad, V.~G.~M.
  Naidu, D.~Bahadur, R.~Srivastava, Polymeric core--shell combinatorial
  nanomedicine for synergistic anticancer therapy, ACS omega 4~(22) (2019)
  19614--19622.

\bibitem{yang2023phase}
G.~Yang, Y.~Liu, S.~Jin, Y.~Hui, X.~Wang, L.~Xu, D.~Chen, D.~Weitz, C.-X. Zhao,
  Phase separation-induced nanoprecipitation for making polymer nanoparticles
  with high drug loading: Special collection: Distinguished australian
  researchers, Aggregate 4~(2) (2023) e314.\\
\textcolor{blue}{* This paper introduces a salt concentration screening method that induces liquid-liquid phase separation to delay the precipitation times of drugs and polymers, facilitating their co-precipitation and enabling the fabrication of polymer nanoparticles with high drug loading.                     }


\bibitem{eltaib2025polymeric}
L.~Eltaib, Polymeric nanoparticles in targeted drug delivery: Unveiling the
  impact of polymer characterization and fabrication, Polymers 17~(7) (2025)
  833.

\bibitem{yin2025novel}
L.~Yin, Q.~Zhang, W.~Zhang, W.~Qiu, C.~Lin, J.~Wang, Novel cellulose
  nanospheres with tunable sizes prepared from nitrocellulose through
  nanoprecipitation, Journal of Applied Polymer Science (2025) e57343.\\
\textcolor{blue}{** This paper reported that salt species can act chemically and play a dual role in nanoprecipitation: HS$^-$ from NaHS both chemically denitrates nitrocellulose and, via ionic strength, promotes salting-out, lowering solubility and triggering nucleation.  }


\bibitem{ding2023aqueous}
D.~Ding, L.~Gong, M.~Li, X.~Cheng, H.~Peng, Z.~Zhang, S.~Wang, X.~Yan, Aqueous
  nanoprecipitation for programmable fabrication of versatile biopolymer
  nanoparticles, Green Chemistry 25~(10) (2023) 4004--4012.

\bibitem{combes2022protein}
A.~Combes, K.-N. Tang, A.~S. Klymchenko, A.~Reisch, Protein-like particles
  through nanoprecipitation of mixtures of polymers of opposite charge, Journal
  of Colloid and Interface Science 607 (2022) 1786--1795.\\


\bibitem{ristroph2019hydrophobic}
K.~D. Ristroph, R.~K. Prud'homme, Hydrophobic ion pairing: encapsulating small
  molecules, peptides, and proteins into nanocarriers, Nanoscale Advances
  1~(11) (2019) 4207--4237.

\bibitem{lu2018encapsulation}
H.~D. Lu, K.~D. Ristroph, E.~L. Dobrijevic, J.~Feng, S.~A. McManus, Y.~Zhang,
  W.~D. Mulhearn, H.~Ramachandruni, A.~Patel, R.~K. Prud’homme, Encapsulation
  of oz439 into nanoparticles for supersaturated drug release in oral malaria
  therapy, ACS Infectious Diseases 4~(6) (2018) 970--979.

\bibitem{ristroph2021highly}
K.~D. Ristroph, P.~Rummaneethorn, B.~Johnson-Weaver, H.~Staats, R.~K.
  Prud'homme, Highly-loaded protein nanocarriers prepared by flash
  nanoprecipitation with hydrophobic ion pairing, International Journal of
  Pharmaceutics 601 (2021) 120397.

\bibitem{fan2021continuous}
W.~Fan, F.~Zhao, J.~Dou, X.~Guo, Continuous preparation of dual-mode
  luminescent lapo4: Tb3+, yb3+ nanoparticles by reactive flash
  nanoprecipitation, Chemical Engineering Science 242 (2021) 116734.

\bibitem{maiocchi2025development}
S.~Maiocchi, E.~E. Burnham, A.~Cartaya, V.~Lisi, N.~Buechler, R.~Pollard,
  D.~Babaki, W.~Bergmeier, N.~M. Pinkerton, E.~M. Bahnson, Development of
  dnase-1 loaded polymeric nanoparticles synthesized by inverse flash
  nanoprecipitation for neutrophil-mediated drug delivery to in vitro thrombi,
  Advanced Healthcare Materials 14~(15) (2025) 2404584.\\
\textcolor{blue}{* This paper reports that the addition of ionic compounds such as spermine and $\rm CaCl_2$ enables ionic cross-linking and pH modulation within the inverse nanocarrier core during inverse flash nanoprecipitation, significantly enhancing the stability, encapsulation efficiency, and activity preservation of DNase-1-loaded polymeric nanoparticles. }


\bibitem{condello2025two}
A.~Condello, E.~Piacentini, V.~Sebastian, L.~Giorno, Two-step and one-step
  approach for the in situ synthesis of palladium nanosheets on zein
  nanoparticle surface using membrane nanoprecipitation, Emergent Materials
  (2025) 1--13.

\bibitem{ruan2024polymeric}
B.~Ruan, Z.~Zheng, A.~B. Kayitmazer, A.~Ahmad, N.~Ramzan, M.~S. Rafique,
  J.~Wang, Y.~Xu, Polymeric ph-responsive metal-supramolecular nanoparticles
  for synergistic chemo-photothermal therapy, Langmuir 40~(32) (2024)
  16813--16823.

\bibitem{shamsipur2023phototriggered}
M.~Shamsipur, A.~Ghavidast, A.~Pashabadi, Phototriggered structures: Latest
  advances in biomedical applications, Acta Pharmaceutica Sinica B 13~(7)
  (2023) 2844--2876.

\bibitem{cheng2024template}
X.~Cheng, S.~Wang, J.~Bernard, F.~Ganachaud, X.~Yan, Template-free
  nanostructured particle growth via a one-pot continuous gradient
  nanoprecipitation, Aggregate 5~(1) (2024) e427.

\bibitem{kang2023photo}
X.~Kang, Y.~Zhang, J.~Song, L.~Wang, W.~Li, J.~Qi, B.~Z. Tang, A
  photo-triggered self-accelerated nanoplatform for multifunctional
  image-guided combination cancer immunotherapy, Nature Communications 14~(1)
  (2023) 5216.

\bibitem{patel2024enhancing}
A.~M. Patel, S.~R. Patel, Enhancing solubility of polymer-loaded febuxostat
  through ultrasound-assisted microfluidic antisolvent nanoprecipitation:
  Optimization using box-behnken design, Chemical Engineering and
  Processing-Process Intensification 201 (2024) 109802.

\bibitem{behera20243d}
R.~Behera, S.~R. Patel, 3d-printed microfluidics under ultrasonic cooling
  crystallization for nano-precipitation of ezetimibe: Effect of process
  parameters and pbd approach, Chemical Engineering and Processing-Process
  Intensification 204 (2024) 109921.

\bibitem{song2023near}
J.~Song, X.~Kang, L.~Wang, D.~Ding, D.~Kong, W.~Li, J.~Qi, Near-infrared-ii
  photoacoustic imaging and photo-triggered synergistic treatment of thrombosis
  via fibrin-specific homopolymer nanoparticles, Nature Communications 14~(1)
  (2023) 6881.

\bibitem{di2023machine}
V.~Di~Francesco, D.~P. Boso, T.~L. Moore, B.~A. Schrefler, P.~Decuzzi, Machine
  learning instructed microfluidic synthesis of curcumin-loaded liposomes,
  Biomedical Microdevices 25~(3) (2023) 29.

\end{thebibliography}

\end{sloppypar}

\end{document}